New Materials for Thin-Film Phase-Plates

# DIPLOMA-THESIS

submitted by

Leonard Frederic Henrichs

2011

**Institut für physikalische und theoretische Chemie**
**Fachbereich Chemie**
**Mathematisch-Naturwissenschaftliche Fakultät**
**Rheinische Friedrich-Wilhelms-Universität Bonn**

First examiner: Prof. Dr. K. Wandelt, Institute for Physical and Theoretical Chemistry, University of Bonn
Second examiner: Prof. Dr. U. Kubitschek, Institute for Physical and Theoretical Chemistry, University of Bonn

Leonard Frederic Henrichs

Eduard-Pflüger-Strasse 58

53113 Bonn

Date of birth: 5th June 1987, Siegen (Germany)

**Eidesstattliche Erklärung:**

Ich versichere hiermit, die Arbeit selbstständig angefertigt zu haben und alle benutzten Hilfsmittel unter Angabe der Literaturstellen kenntlich gemacht zu haben.

**Declaration in lieu of an oath:**

I hereby declare that the work submitted is my own and that all passages and ideas that are not mine have been properly cited.

Bonn, December 2011

Für Juliane & Jesper

# Abstract

Previous Zernike phase-plates, for transmission-electron-microscopy (TEM) have always been made out of carbon. A critical constraint on those phase-plates is their fast degradation during use. Here, a Zernike phase-plate consisting out of metal is produced and tested for the first time. For the new aluminium phase-plate, no charging like for the carbon phase-plate is observed. Under conditions of use, the stability of the aluminium phase-plate, is excellent. For both materials, a distortion in the image's Fourier-transformation is observed, when inserting the thin-film into the beam at the back-focal-plane. This distortion is a negative, spatial-frequency dependent phase-shift, that is not equal to a focus-change. A phase-shift and an increase of contrast, is observed for the tested phase-plates.

Different types of self-sustaining metal-films are produced with a technique, invented by Janbroers et al. All metal-films, produced at room-temperature, exhibit a polycrystalline structure, while carbon-films are amorphous. However, a high crystallinity is unfavored, causing distortions in the image and an increase of electron-scattering in the phase-plate, due to diffraction. Using cryo-deposition, metal-films are obtained that combine an amorphous structure with beneficial properties of metals such as lower charging and prolonged durability. Investigations with TEM, show a great potential of those cryo-films as material for Zernike phase-plates.

Thickness-measurements are successfully carried out with atomic-force-mycroscopy at step-edges between the thin-film and its substrate. For this, a technique for the preparation of very narrow step-edges is developed here.

# Contents

# 1 Introduction

## 1.1 Benefits of Transmission-Electron-Microscopy in Life-Sciences

Today, an important tasks in life-sciences is the determination of the sub-cellular structure of the body. Knowing biological-structures on molecular level is important for better understanding processes inside the cell. A good example for this is the Alzheimer's disease, which is connected to the formation of amyloid-plaques [1] and abnormal tau-proteins [2] in the brain. The knowledge about their structure, occurrence and effects on our brain is most important for possible treatments. This knowledge can be obtained using microscopic techniques. Fluorescence-microscopy and related techniques such as confocal laser-scanning- [3] and two-photon excitation-microscopy [4] are nowadays widely used methods to image whole cells or clusters of cells. However, the resolution of those techniques, is limited to the region of tens of $nm$.

The only microscopic technique capable of resolving biological-structures below $20\,nm$ is transmission-electron microscopy (TEM). In addition, electron-tomography has become a powerful method, to obtain 3D-structural information of sub-cellular structures. Several reviews and articles have been published on this topic in the past years [5, 6, 7, 8].

Preparing biological samples for TEM is challenging, because of the water that is natively present in those samples. Water would evaporate in the high-vacuum of a TEM, whereby the sample would be destroyed. One way to prepare biological samples, is to fixate them by replacing the contained water with a polymer. However, by treating the sample with chemicals, the in-vivo structure, usually can not be conserved.

The ultimate preparation-method for biological TEM-samples is rapid freezing. By this, the water solidifies amorphously and the in-vivo structure of the cell is best preserved. Using such vitrified samples for cryo-TEM, it was possible to determine biological-structures at a resolution of $4\,\text{Å}$ [9].

However, there are certain problems concerning the contrast of biological samples in TEM.

## 1.2 Problem of Contrast in Cryo-TEM

For biological samples that consist mostly of elements with low atomic-number (H, C, O, N), the contrast in TEM is poor due to the weak interaction between those elements and the high energetic electrons used in conventional TEM. This problem is usually reduced by staining samples with heavy-metal salts such as uranyl-acetate, which has the disadvantage of chemical-artifacts [10]. In addition, vitrified biological samples exhibit a high sensitivity to the electron-beam and thus to possible changes of the in-vivo structure of the sample. In cryo-EM, staining is not possible at all. Staining-solutions are water based and would immediately freeze on the sample surface, rendering the thin film useless. Hence, only low doses - 30 electrons per $\text{Å}^2$ as a rule of thumb - can be applied to samples without inflicting damage in cryo-electron microscopy. In cryo-ET a large number of images has to be collected from the same sample-position in order to obtain enough data for 3D-reconstruction. Usual electron doses in cryo-ET for one image are 1-2 electrons per $\text{Å}^2$. The resulting data are normally just above the noise level of the detector. In addition, obtaining pictures with higher contrast, requires longer acquisition-times and a number of images with different underfocus-settings, which in turn increases the applied electron-dose [11]. Therefore higher contrast for biological samples, without staining and fixation is needed. A solution for this problem is to enhance the so-called phase-contrast by using phase-plates.

## 1.3 Phase-Contrast: Solution for the Problem of Low Contrast

The concept of phase-contrast using phase-plates was developed by Frits Zernike for conventional light-microscopy in 1932. He received the nobel-price for his works in 1953. In 1947 Hans Boersch proposed two designs of a phase-plate for TEMs [12] shown in figure 1.1, which is from Boersch's article.

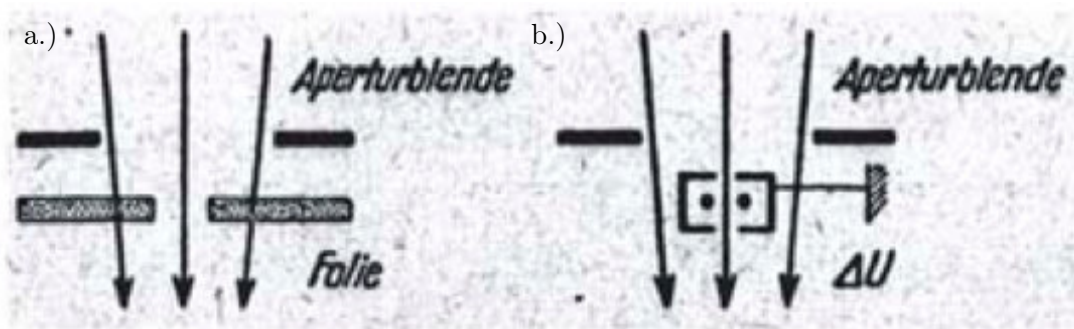

**Figure 1.1:** Phase-plates for TEM proposed by Hans Boersch. a.) Zernike phase-plate; b.) electrostatic Boersch phase-plate [12].

One phase-plate is in analogy to Zernike's phase-plate which works through a thin-foil. The other works through an electrostatic potential. Accordingly, the two different types of phase-plates are called Zernike-, and Boersch phase-plate.

Unfortunately, the implementation of a phase-plate in a TEM is a much bigger technical challenge, compared to light microscopy (like is every other aspect of the two techniques). This is mainly due to the needed micrometer-dimension of phase-plate for a TEM. First attempts to implement the proposed phase-plates were done in 1958 [13]. However, only a hole of $20\,\mu m$ diameter could be produced at the time. This was not nearly small enough to successfully use phase contrast in a TEM. The reason for this will be given later in section 2.1.5. Other efforts to implement phase-plates in TEM were done in the 1970s [14, 15]. A thin wire-type electrostatic phase-plate and an improved Zernike phase-plate were developed, but yielded only minimal improvement in contrast.

In 1999 Kuniaki Nagayama published a first theoretical paper on the topic, after a period of approx. 20 years of silence [16]. The breakthrough came in 2001 when a big gain in contrast was reported by Radostin Danev, and Kuniaki Nagayama with their improved carbon Zernike phase-plate [17] with a hole diameter of $1\,\mu m$. Since then, further investigations about Zernike phase-contrast [18, 19, 20] as well as other types of phase-plates like a half-plane Hilbert phase-plate [21] or an electrostatic Boersch phase-plate [22, 23, 10] have been published. The number of publications on the application of phase-contrast in cryo-EM [11, 24, 25], and of reviews [26, 27] about the topic, has tremendously increased since the early 2000s.

The effect of a Zernike phase-plate is illustrated in figure 1.2.

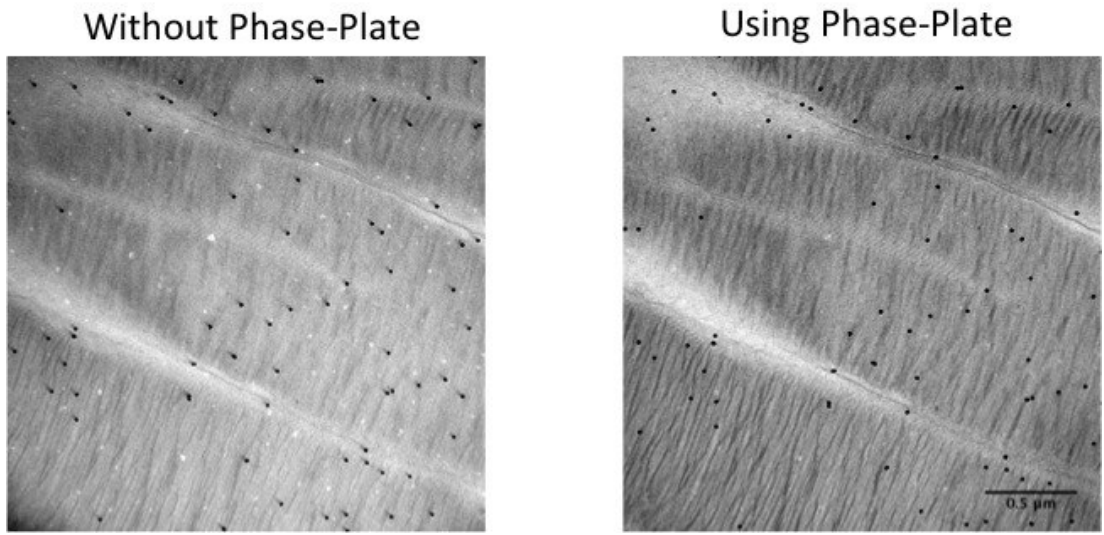

**Figure 1.2:** Bright-field TEM-micrographs of mouse-retina without using a phase-plate (left) and using a phase-plate (right). Courtesy of Monika Fritz, center of research "caesar".

An increase in contrast of this vitrified sample (mouse retina) is clearly visible. Details are visible much better in the right panel.

Phase-contrast with phase-plates is currently object to extensive research of several groups all over the world.

## 1.4 Aims of This Thesis

This diploma-thesis aims at making a contribution in the fast developing field of phase-contrast for TEM. First of all, carbon Zernike phase-plates comparable to those made by Danev et al. shall be reproduced in order to have a benchmark for the following experiments. To overcome problems of the carbon based phase-plates, mainly their fast degradation while using, new materials for phase-plates will be tested. In a first step, suitable metals will be selected. Subsequently, experiments on the stability of new phase-plates when exposed to the electron beam are to be carried out. Another goal is the characterization of the produced phase-plates. Here, properties of interest are: thickness, crystalline structure and nano-morphology.

The following thesis, is divided into four chapters. Starting with a chapter about the general theory, it proceeds with an individual chapter for each different type of Zernike phase-plate produced during the work on this thesis. Those different types were: carbon-, metal-, and cryo-metal phase-plates. The preparation, utilization as phase-plate, characterization and discussion of results is presented for each type of phase-plate separately in the according chapters.

# 2 General Theory

*Before treating the principles of phase-contrast generated by a phase-plate, the following chapter will first provide an overview about the processes of image-formation in TEM. Afterwards, an introduction into the topic of phase-plates will be given.*

## 2.1 Image-Formation and Phase-Contrast

In general, image-contrast is due to intensity variations of neighboring regions in an image, e.g. pixels on a CCD-camera. We can distinguish between two types of contrast: amplitude- and phase-contrast.

Amplitude-contrast originates from a "loss" of electrons in the TEM-column, caused by scattering-processes inside the specimen. In conventional TEM, electrons have normally energies between $80\ldots300\,keV$ and are hence not absorbed by the thin sample $(10\ldots100\,nm)$. Usually, samples are not homogeneous but show variations in e.g. mass-density (Z-number) or crystallographic orientation. Denser areas in the sample scatter electrons stronger, resulting in a higher "loss" of electrons; these areas appear darker in the image. In conventional TEM-micrographs, e.g. of inorganic materials consisting of high-Z elements, image-contrast is almost completely due to amplitude-contrast.

Phase-contrast results from a phase-difference of parts of the electron-wave. Here, contrast is produced by constructive and destructive interference of phase-shifted and unchanged parts of the electron-wave in the image. A more detailed description of phase-contrast is given in the following subsections.

### 2.1.1 Weak-Phase/Weak-Amplitude Approximation

In a TEM a sample is illuminated by a parallel $e^-$-beam which corresponds to a planar electron-wave. The amplitude of the specimen exit-wave is a function of the radial-coordinate $r_0(x_0, y_0)$ in the specimen-plane [22] and can be expressed as:

$$\psi_{exit}(r_0) = [1 - A(r_0)] \cdot exp(-i\eta(r_0)) \tag{2.1}$$

$A(r_0)$ stands for the reduction of amplitude through higher-angle scattering processes and hence for amplitude-contrast. $\eta$ is determined by the one-dimensional projection of the electrostatical-potential $\delta$ of the specimen with a uniform thickness $t$ at $r_0$. Furthermore, $\eta$ depends on the electron-wavelength $\lambda$ and the acceleration-voltage $U_0$. $\eta$ can be interpreted as the distortion of the wave-front caused by a phase-shift of parts of the electron-wave [22]:

$$\eta(r_0) = \frac{\pi}{\lambda U_0} \cdot \frac{2(m_0 c^2 + eU_0)}{(2m_0 c^2 + eU_0)} \int_{z_0+\frac{t}{2}}^{z_0-\frac{t}{2}} \delta(r_0, z) dz \tag{2.2}$$

$m_0$ is the electron rest mass and $e$ the electron-charge. We can develop the term $exp(-i\eta(r_0))$ in equation 2.1 as a Taylor-series:

$$exp(-i\eta(r_0)) = 1 - i\eta(r_0) + \frac{\eta(r_0)^2}{2}... \tag{2.3}$$

In a thin sample composed of low-z-elements e.g. a vitrified biological sample, the change of the phase is small. Such samples are called weak-phase objects. Here all higher terms of the Taylor-Series can be neglected, which is known as the weak-phase-approximation. Products of $A(r_0)$ and $i\eta(r_0)$ can be neglected as well, because both factors are small. This yields if we substitute in equation 2.1:

$$\psi_{exit}(r_0) = 1 - A(r_0) - i\eta(r_0) + i\eta(r_0)A(r_0) \approx 1 - A(r_0) - i\eta(r_0) \tag{2.4}$$

The distortion of the wave-front due to a weak-phase object - or just phase-object - is illustrated in figure 2.1.

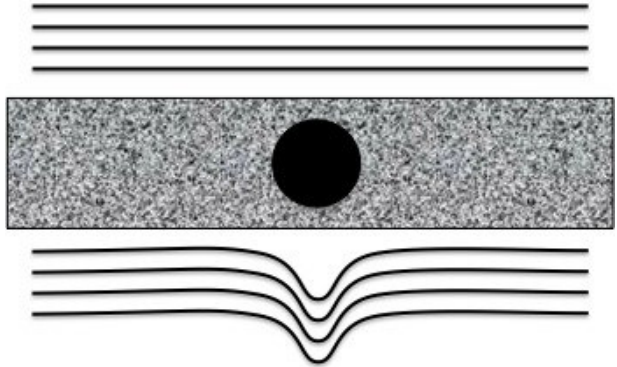

**Figure 2.1:** A weak-phase object depicted as a globule, is embedded in another medium (e.g. amorphous ice). The phase of the incoming planar wave is locally shifted, where it has intersected the object. Hence the exit-wave front is deformed.

Furthermore, the modulation of the amplitude is much lower than that of the phase for these kind of samples, thus we can neglect the term $A(r_0)$ in comparison to $i\eta(r_0)$ (weak-amplitude-approximation).

We can now identify the terms resulting from the combined weak-phase/weak-amplitude approximation, as contributions from the scattered parts $\psi_{sc}$ and unscattered parts of the electron-wave $\psi_0$:

$$\psi_0 = 1 \tag{2.5}$$

$$\psi_{sc} = -i\eta(r_0) \tag{2.6}$$

Under small scattering-angles, the phase of scattered parts of the electron-wave are shifted about $90°$ or $\pi/2$ relative to the unscattered [28] which is represented in equation 2.4 by the imaginary factor $i$.

The intensity resulting from an electron-wave is defined as the square of the absolute value of the amplitude [28]:

$$I = |\psi|^2 \tag{2.7}$$

and therefore depends on the vector that results from plotting the complex amplitude of the electron-wave. This is shown in figure 2.2.

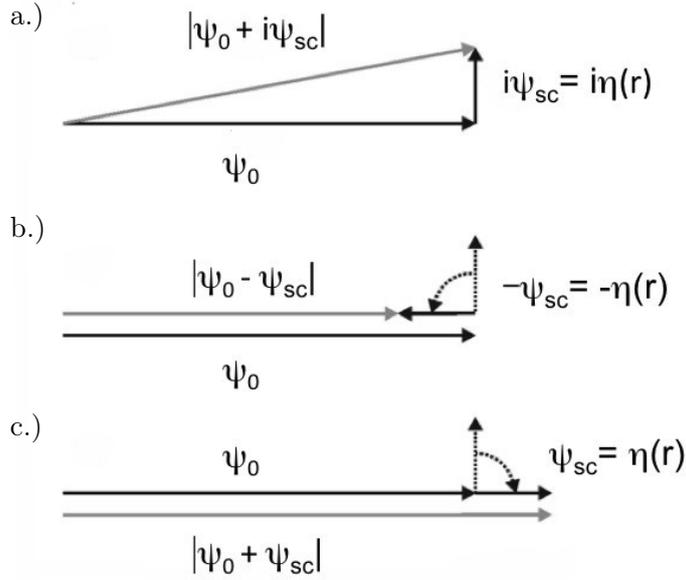

**Figure 2.2:** The interference of scattered with unscattered parts of the electron-wave is imaged schematically. a.) In weak-phase objects, scattered parts of the electron-wave are shifted by $\pi/2$ respectively to the unscattered parts. The resulting intensity which corresponds to the length of the vector, does not differ significantly from the background, therefore the phase-contrast is negligible. b.) Positive phase-contrast: By a shift of further $\pi/2$, the vector of the scattered part of the electron-wave gets parallel to the unscattered, and maximum phase-contrast is achieved. c.) Negative phase-contrast: The scattered part is shifted by $-\pi/2$ or $3/2\pi$[23].

Panel a.) shows the "normal" scenario: If the amplitude of the scattered wave $\psi_{sc}$ is small in comparison to that of the unscattered $\psi_0$, the sum $\psi_0 + i\psi_{sc}$ does not differ significantly from $\psi_0$. This is because $\psi_{sc}$ is perpendicular to $\psi_0$ which is expressed by the complex factor $i$. Panel b.) depicts the vector of the amplitude when the phase of the scattered wave is additionally shifted by $\pi/2$. The resulting amplitude $\psi_0 - \psi_{sc}$ differs much more from $\psi_0$ than without phase-shift, because $\psi_{sc}$ is now parallel to $\psi_0$. This scenario is called positive phase-contrast. A phase change of $-\pi/2$ or $3/2\pi$ results in an intensity $I = |\psi_0 + \psi_{sc}|^2$ (negative phase-contrast) which is shown in figure 2.2 c.).

Now, that we have learned about phase-contrast in context of the weak-phase/weak-amplitude approximation, the next step is quantifying of the image-contrast.

## 2.1.2 Contrast-Transfer Function

A very useful approach to describe the contrast-properties of an optical-system, is the contrast-transfer function. It delivers a value of contrast-transfer, as a function of the reciprocal wave-vector $k$ of the image's Fourier-spectrum.

For a specimen with a single spatial-frequency, the contrast-transfer function of the amplitude ($aCTF$) and the phase-structure ($pCTF$) can be generally written as [28]:

$$aCTF(k) = -\cos(\gamma(k)) \tag{2.8}$$

$$pCTF(k) = \sin(\gamma(k)) \tag{2.9}$$

with the $CTF$-phase-shift $\gamma(k)$ :

$$\gamma(k) = 2\pi(-\frac{1}{2}\Delta z\lambda k^2 + \frac{1}{4}C_s\lambda^3 k^4) \tag{2.10}$$

here $\Delta z$ denotes the defocus and $C_s$ is the spherical-aberration constant of the microscope's objective-lens. In case of a phase-object, one can neglect the amplitude-structure so that the specimen is described only by its phase contrast-transfer function which will be corresponded to as $CTF$.

In figure 2.3, the $CTF$ for an imaging system with an accelerating-voltage of $200\,kV$, a defocus of $-10\,nm$, and a spherical-aberration of $2.2\,mm$ is plotted.

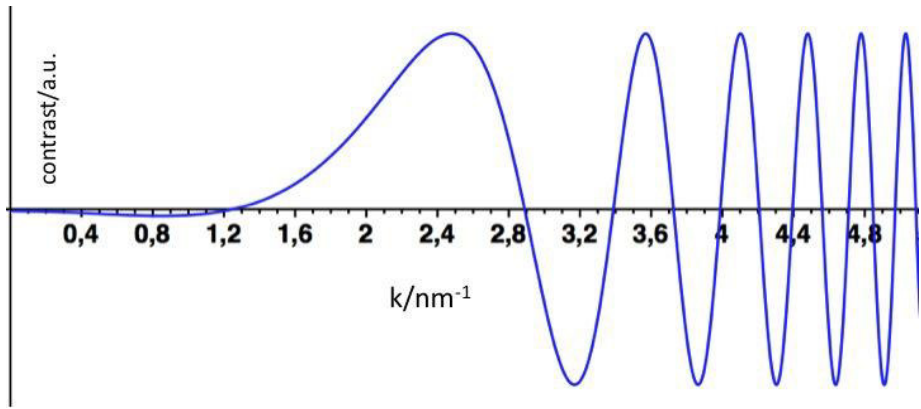

**Figure 2.3:** Simulated $CTF$ for an acceleration-voltage of $200\,kV$, a defocus of $-10\,nm$ and a spherical-aberration of $2.2\,mm$

Here, an idealized $CTF$ is displayed. In reality, it is damped exponentially for increasing spatial-frequency, due to aberrations. Because the general shape of the CTF is not affected by the damping, it is not shown here. The conventional $CTF$ is a sine-function, therefore no or very low contrast is transferred, in the region from $0\ldots1.6\,nm^{-1}$. This corresponds to objects larger than $0.63\,nm$. Phase-objects of this size are invisible in conventional TEM. This is especially problematic because objects in biological samples, like membranes, cell-organella or proteins, have a size of $3\ldots100\,nm$. This issue becomes evident in figure 2.4. Panel a.) shows model simulations of ribosomes which are approx. $25\,nm$ large. When ice embedding and thermal noise is added to the model simulation(panel b.), no structures can be seen near the focus. In the following subsections, solutions for this problem will be discussed.

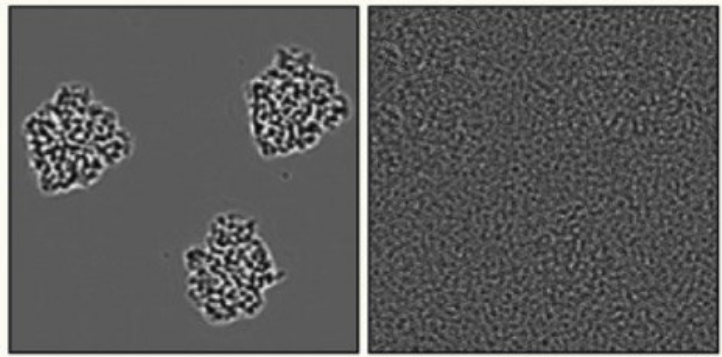

**Figure 2.4:** Simulations of TEM-micrographs of Ribosomes with $200\,nm$ underfocus: a.) Under ideal conditions; b.) With ice embedding and 10% thermal noise [22].

### 2.1.3 Phase-Contrast

Phase-contrast can be enhanced by defocusing. In figure 2.5 a $CTF$ for an underfocus of $500\,nm$ is shown.

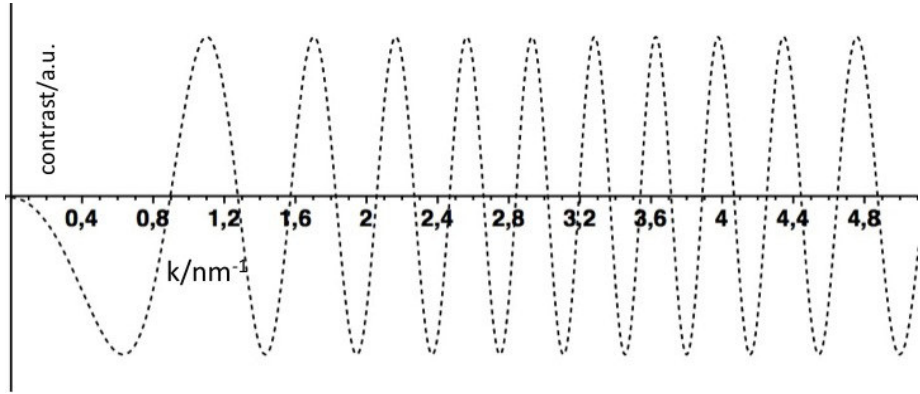

**Figure 2.5:** Simulated $CTF$ for an acceleration-voltage of $200\,kV$, a defocus of $-500\,nm$ and a spherical-aberration of $2.2\,mm$.

By defocusing the $CTF$ is compressed and the contrast transfer in the region of $0.4\ldots0.8\,nm^{-1}$ becomes large. This means that objects with sizes ranging from $1.25\ldots2.5\,nm$ can be seen under these conditions. Figure 2.6 shows the same simulations as in figure 2.4 except that a defocus of $-1000\,nm$ (underfocus) is applied here. With underfocus, ribosomes can be distinguished from the background (panel b.) under realistic conditions.

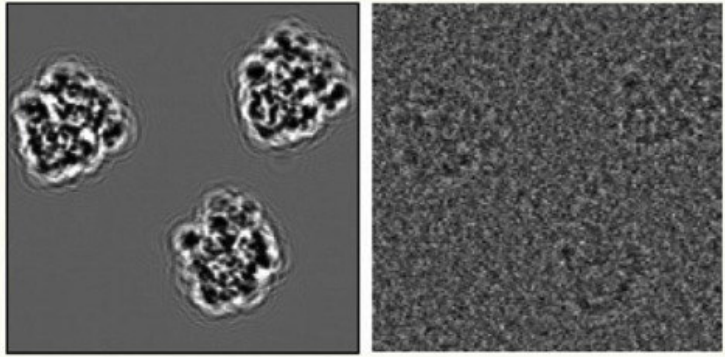

**Figure 2.6:** Simulations of TEM-micrographs of Ribosomes with $1000\,nm$ underfocus: a.) Under ideal conditions; b.) With ice embedding and 10% thermal noise [22].

However, through the high periodicity of the $CTF$, the image contains transfer-gaps for certain spatial-frequencies and phase-flipping. This causes blurring and undesired effects like halos, and complicates the image-interpretation and decreases the resolution, as can be seen in panel a.).
Besides defocusing, the $CTF$ can be influenced by introducing a phase-plate into the system. It induces a phase-contrast phase-shift $\phi$ to the $CTF$ [17]:

$$CTF(k) = \sin[\gamma(k) + \phi] \tag{2.11}$$

with

$$\phi = -\pi \frac{2tV_0(m_0c^2 + eU_0)}{\lambda U_0(2m_0c^2 + eU_0)} \tag{2.12}$$

By applying a phase-shift of -$\pi/2$ to the $CTF$, it is transformed from a sine-function into a cosine-function. This ideal phase-contrast-$CTF$ is shown in figure 2.7.

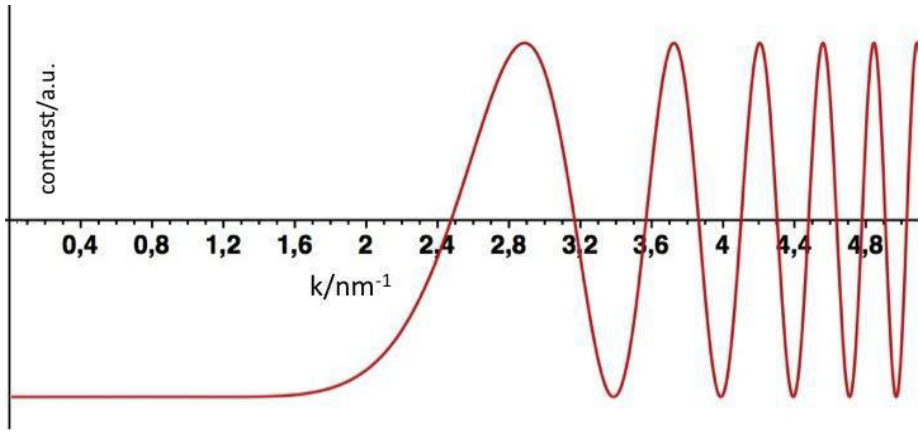

**Figure 2.7:** Simulation of an ideal -$\pi/2$ shifted phase-contrast-$CTF$ at Gaussian-focus for an acceleration-voltage of $200\,kV$ and a spherical-aberration of $2.2\,mm$.

Due to the fact that the phase-contrast-$CTF$ is a cosine-function that starts with a maximum value at low spatial-frequencies, the ideal imaging condition in terms of contrast, is at focus. Here a wide range of spatial-frequencies have a high uniform contrast-transfer, and of course the image has the highest resolution. Objects ranging from about $0.5\,nm$ to infinite size should be imaged with high contrast, by using a phase-plate.
However, in reality there are limitations to the phase-contrast-$CTF$ due to the design of phase-plates. The working-principle, design and limitations of different types of phase-plates, are described in section 2.1.4 and further subsections.

### 2.1.4 Phase-Plates

For better understanding of this subsection, a very basic buildup and optical-path of a TEM is presented schematically in figure 2.8. The electron-gun, e.g. a thermoionic- or a field-emission-gun, is positioned at the top of an evacuated column. A set of electromagnetic condensor-lenses deflect the electrons, so that the specimen is illuminated homogeneously by a parallel electron-beam. After the beam has transmitted the specimen, the objective-lens forms the image. In the objective-lens' focal-plane (back-focal-plane) a diffraction pattern is formed. All unscattered electrons are focused in one point on the optical axis, while scattered electrons are focused in off-axial points, corresponding to their scattering-angle. The first real-space image is formed, where scattered and unscattered electrons coincide again. The electrons propagate to a system of projector-lenses, magnifying the image, and finally projecting it onto an imaging-device e.g. a fluorescent-screen or a CCD-chip. By changing the focal-length of the first intermediate projector-lens, the diffraction-image can also be recorded.

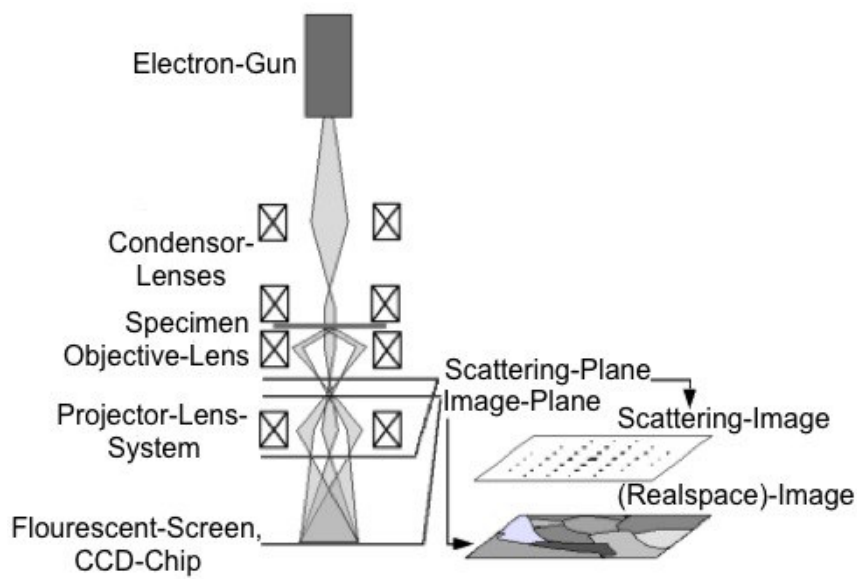

**Figure 2.8:** Schematic assembly and optical-path of a TEM [29].

A phase-plate is usually inserted in the back-focal-plane where scattered and unscattered electrons are separated. There are two different types of phase-plates: matter-less phase-plates which utilize an electrostatic potential, and phase-plates working through interaction of the electron-beam and matter. Zernike phase-plates belong to the latter type. Here, the phase of the scattered part of the electron-wave is delayed by the mean inner-potential of the material. The mean inner-potential describes how strong matter interacts with electrons. Detailed information about it will be given later in subsection 2.1.6. In figure 2.9 the principle of different phase-plates in TEM are illustrated.

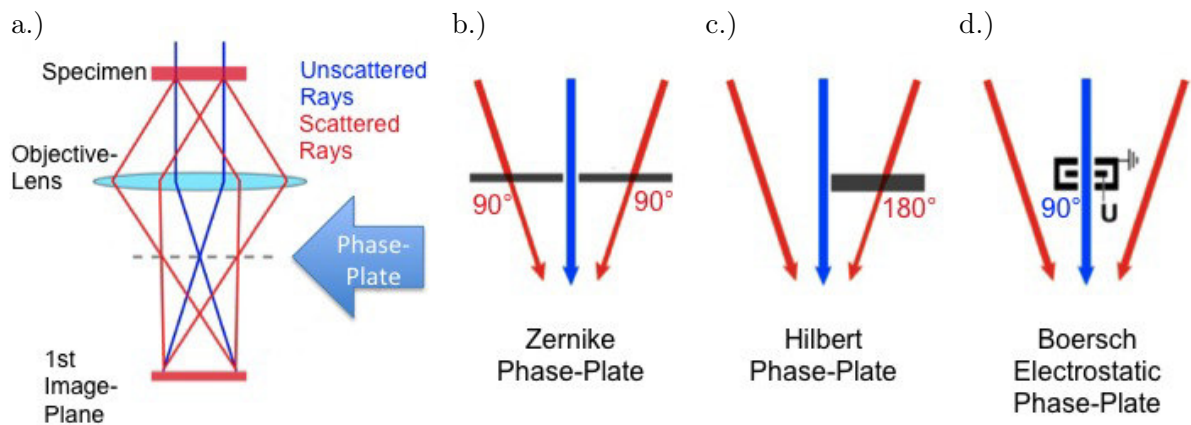

**Figure 2.9:** a.) Schematic buildup and optical path of a TEM in the region of the back-focal-plane; Sideviews in the back-focal-plane of: b.) Zernike phase-plate; c.) Hilbert phase-plate; d.) Boersch phase-plate [22].

Panel a.), shows the optical path of a TEM from the specimen-plane till the plane where the first image is formed. The blue arrow marks the back-focal-plane, where a phase-plate can be inserted. Here the unscattered and the scattered electron-beams are spatially-seperated.

b.) shows a Zernike phase-plate from a sideview, lying in the back-focal-plane. The unscattered electrons propagate through a central hole in the phase-plate while the scattered electrons are transmitted through the thin-film phase-plate. In an ideal case, the phase-plate has a thickness so that all scattered electrons obtain a phase-shift of $-\pi/2$. Additionally, the central hole should be infinite small, so that only unscattered electrons, lying on the optical axis, could pass through the phase-plate unchanged. This is of course an idealization and can not be realized (see section 2.1.5). Similar to the Zernike phase-plate, the Hilbert phase-plate also uses the inner-potential of matter for a phase-shift (panel c.). In contrast to the round Zernike phase-plate it consists of a half plane, by which half of the scattered electrons obtain a phase-shift of $\pi$ and the other half plus the unscattered electrons remain unchanged. Through a process of image-reconstruction, a similar phase-contrast as for a Zernike phase-plate is obtained [21].

The third type is a matter-less phase-plate, where the phase-delay is generated by an electrostatic-potential. There are several possible designs for such an electrostatic phase-plate but only the original design proposed by Boersch could be successfully realized so far [22, 23, 10]. This miniature electrostatic-lens is called Boersch phase-plate and is shown schematically in panel d.). Here the unscattered electrons take course through the center of the miniature electrostatic lens. The potential of the lens can be adjusted in a way, that the unscattered electrons are shifted about $-\pi/2$ relative to the scattered electrons. This is an important difference to the matter-containing phase-plates, where the scattered electrons are shifted in phase.

Other types of electrostatic phase-plates are e.g. the anamorphotic- and the Zach phase-plate [23]. Ideally, all scattered electrons should pass by the Boersch phase-plate, and just unscattered electrons should go through the einzel-lens. From a technical point of view, it is very difficult to realize such lenses. An example of a Boersch phase-plate which has been realized by Barton et al. is shown in figure 2.10.

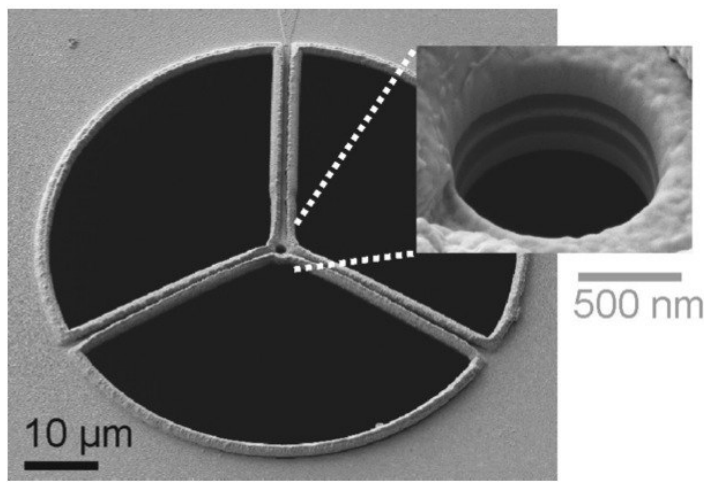

**Figure 2.10:** Boersch phase-plate produced by Schultheis et al. Diameter inner ring: $1.2\,\mu m$; Diameter of outer-ring: $3.5\,\mu m$ [23]

Since only Zernike type phase-plates were investigated during the work for this thesis, only these will be described in detail.

### 2.1.5 Zernike Phase-Plate and Cut-On Frequency

The phase-shift $\phi$ of a Zernike phase-plate depends on the mean inner-potential $V_0$ of its material. The phase-contrast-mechanism of a Zernike phase-plate is based on the existence of a small central hole, through that unscattered electrons can pass unchanged. The size of this hole is connected to the so-called cut-on frequency $k_h$, which is the lowest spatial-frequency upon the phase-shift is applied. It depends on the radius $r_h$ of the central hole and is given by [17]:

$$k_h = \frac{r_h}{f\lambda} \tag{2.13}$$

Here, f is the focal-length of the objective-lens. Phase-plates produced for this thesis, always had a nominal diameter of the central hole of $1\,\mu m$. For a $200\,kV$ TEM with a focal length of its objective-lens of $3.0\,mm$, the cut-on frequency is $1/15.06 = 0.066\,nm^{-1}$. This cut-on frequency is included in the phase-contrast-$CTF$ that is depicted in figure 2.11. By the limitation of the cut-on frequency, phase-objects larger than $15.06\,nm$ are invisible. However, this is still enough to image details of biological samples such as membranes, macromolecules or cell-organelles, which have usually dimensions of a few to ten nanometers.

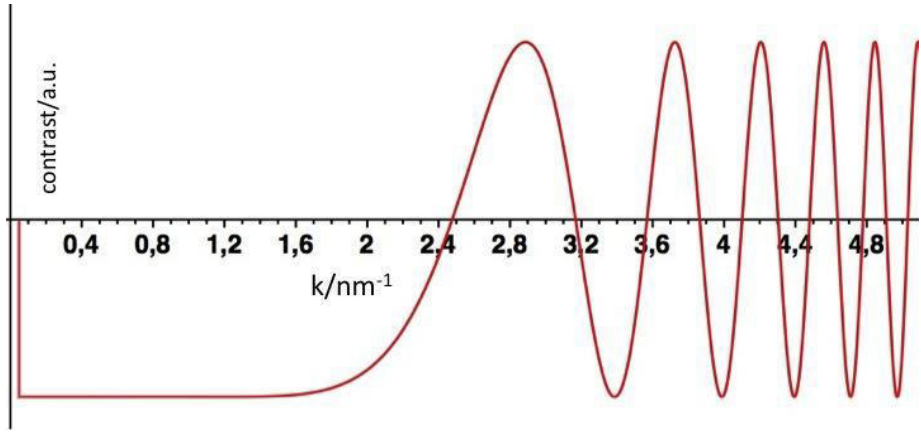

**Figure 2.11:** Simulation of an -$\pi/2$ shifted phase-contrast-$CTF$ with a cut-on frequency of $1/15.06\,nm^{-1}$ at Gaussian-focus for an acceleration-voltage of $200\,kV$ and a spherical-aberration of $2.2\,mm$.

The cut-on frequency can be improved further, by decreasing the size of the central hole, as long as the unscattered beam fully fits inside. In TEMs with a field-emission gun, hole diameters down to $300\,nm$ can be realized. For thermoionic cathodes, the beam diameter is typically in the range of $1\,\mu m$.

In case of the Boersch phase-plate, the outer radius of the einzel-lens has to be taken for calculating the cut-on frequency because of the shadowing-effect of the ring-electrode. Boersch phase-plates produced by Barton and Schultheis had an outer diameter of about $3.5\,\mu m$ (see figure 2.10) limiting the cut-on frequency to $1/4.3\,nm^{-1}$. This is less than a third of the Zernike phase-plates used for this work. The same holds true for the upper limit of the size of phase-objects visible. This is an important advantage of Zernike phase-plates in comparison to Boersch phase-plates.

A disadvantage of Zernike phase-plates is, that scattered electrons, already having a small intensity compared to the unscatterd, are further damped by the phase-plate so that the effect of their interference becomes smaller. The opposite would be ideal. A thin-film that would change the phase of unscattered electrons by -$\pi/2$ and at the same time decreases their intensity. This is the case in phase-contrast light-microscopy. Unfortunately, such a phase-plate can not be realized due to technical difficulties.

## 2.1.6 Mean Inner-Potential

As mentioned in subsection 2.1.5 the phase-shift induced by a Zernike phase-plate is proportional to the mean inner-potential $V_0$. In conventional TEM, the acceleration voltage is far above $10\,kV$. Here, electrons of the beam are scattered exclusively by the coulomb-potential of the cores, and not by the electrons of the sample. $V_0$ is then given by [30]:

$$V_0 = \frac{1}{\Omega} \int d^3 r V_C(r) \tag{2.14}$$

with $\Omega$ denoting the unit-cell volume and $V_C(r)$ the coulomb potential of a crystal. $V_0$ is the integral of the coulomb-potential in all three directions, normalized by the unit-cell-volume. In table 2.1.6 values for $V_0$ from the literature are presented for different elements. Because those values differ greatly from each other, more than one is given for each element. Additionally, calculated thicknesses (see equation 2.12) for Zernike phase-plates producing a phase shift of $-\pi/2$ are given on the right for different acceleration-voltages. Values of $V_0$ were taken for the calculation, which were considered the most likely by judging the quality of the underlying data. The big variety of values of $V_0$ is not surprising when considering the variety of possible structures for one element, caused by different conditions of preparation. Schowalter et al. could show, that the mean inner-potential crucially depends on the mass-density [30] and the bulk-structure of a material. Additionally, a dependence of the thickness of a thin-film was reported by Wanner et al. According to their reports, there is a constant, thickness independent contribution to $V_0$, due to a dipole-layer at the surface [31], which results in a higher effective mean inner-potential for thinner films. The same observation was made by Danev et al. [18], who reported a much higher value of $V_0$ for thinner carbon-films with respect to identically prepared thicker films. Additionally, Danev et al. measured different values for $V_0$ for carbon-films with the same thickness, that were prepared differently. Both authors used the same method, to determine the mean inner-potential.

| | | thickness for $-\pi/2$ phase-shift ($nm$) | | |
|---|---|---|---|---|
| element | $V_0\,(V)$ | $120\,kV^e$ | $200\,kV^e$ | $300\,kV^e$ |
| Be | $7.8^a$ | 23.31 | 27.63 | 30.86 |
| C | $7.8^a$, $9.09^{b,d}$, $10.7^c$ | 20.00 | 23.71 | 26.48 |
| Al | $11.9^a$, $12.4^{a,d}$, $13.0^a$ | 14.66 | 17.38 | 19.41 |
| Cu | $20.1^{a,d}$, $23.5^a$ | 9.05 | 10.72 | 11.97 |
| Au | $21.1^{a,d}$, $22.1$-$27^a$ | 8.62 | 10.21 | 11.41 |

[a]taken from reference [28]
[b]taken from reference [30]
[c]taken from reference [31]
[d]value taken for thickness calculation
[e]acceleration-voltage

**Table 2.1:** Values for $V_0$ for different elements and calculated phase-plate thicknesses for different acceleration-voltages.

According to these reports, $V_0$ is highly susceptible on the preparation-conditions affecting the surface- and bulk-structure and also the density of a thin film. It is therefore problematic, to adopt values for $V_0$ from the literature, without taking preparation-conditions into account.

Therefore, $V_0$ should be measured for each different preparation-condition individually, and the thickness of the film should be taken into account. Although it is advisable, no measurements of the mean inner-potential were carried out because the effort would have been too much to accomplish during the works for this thesis.

## 2.2 Thin-Film Formation

Thin-films prepared during this thesis, were made by vapor-deposition. Here atoms or particles from the gas-phase condensate on a substrate to form a thin-film. There are two main types of vapor-deposition: physical vapor-deposition (PVD), and chemical vapor-deposition (CVD). While in PVD, gas particles have the same chemical-composition as the source-material, this is not the case in CVD. Three PVD-techniques commonly used are: thermal- and e$^-$-beam-evaporation and sputtering. In both, thermal- and e$^-$-beam-evaporation, the source-material is evaporated by heating it under high-vacuum. For thermal evaporation, the source is usually heated using electric current, while for e$^-$-beam-evaporation it is heated by an electron-beam. For sputtering, an Ar-plasma is used. Here, Ar-ions that are accelerated towards a target, release atoms or particles from the target into the gas-phase.

Thin films have a wide variety of structure. With epitaxial growth, single-crystalline films can be obtained. It only occurs, when the substrate is crystalline and the mismatch between the crystal-lattices of the substrate and the thin-film material is not too high. On amorphous substrates such as carbon-films, the structure of the formed thin-film can either be polycrystalline, or amorphous. An important parameter in thin-film growth is temperature. If the temperature is high, adatoms (atoms that are confined to a surface) can diffuse on a surface to occupy minimum energy positions [32]. At low temperatures, adatoms can not diffuse on the surface, and the thin-film is amorphous. Thus a crystalline lattice is formed. The morphology of the film depends on the tendency of adatoms to form clusters or continuous layers. Polycrystalline-films are often composed out of clusters, causing grain boundaries in the film.

# 3 Experiments and Results

*A general difficulty about phase-plate preparation, is to produce a self-sustaining thin-film that spans over a supporting structure. Thin-films are usually made by vapor-deposition onto a plain substrate. The critical part in the preparation, is always the removal of the film from the substrate and the subsequent transfer onto a supporting-structure like a TEM-grid or a miniature aperture. This chapter is divided into the following sections: carbon-films, metal-films and cryo-metal-films. This partition represents the differences in the preparation-techniques and the properties of the corresponding films. The preparation-techniques will be described. Furthermore methods used to characterize the thin-films are introduced and results on those experiments will be presented. Finally the application results of the thin films as phase-plates will be shown. A discussion on the experiments and results for each kind of thin-film, will close up each section.*

## 3.1 Carbon-Films

To my knowledge, carbon is the only material, phase-plates are currently made of. Those phase-plates according to Danev et al. were produced, in order to have a comparison to new phase-plates.

### 3.1.1 Preparation

Carbon-film phase-plates were made according to Danev [17, 18]: carbon was deposited on freshly cleaved mica via thermal or $e^-$-beam evaporation. Afterwards, the films were floated on water, picked up with a standard copper TEM-grid, and finally dried on several sheets of filter paper. Deposition of carbon-film via thermal evaporation, was done in a modular high vacuum coating system (MED020, BALTEC AG) equipped with a head for thermal evaporation. Two carbon-rods (Ø6.15 mm, BAL-TIC GmbH, spectrally clean) were used as evaporation-source. The end of one of those rods was trimmed in a lathe, to a cylindrically shaped tip, with a diameter of 1.5 mm and a length of 2.5 mm. This tip had a high and homogeneous electrical resistance across the whole length. The shape of the tip was chosen, in order to obtain stable and well defined conditions during deposition. The trimmed tip was pressed against the other rod with springs.

Evaporation was carried out at pressures of $1 \cdot 10^{-5} \ldots 5 \cdot 10^{-5}\,mbar$, and a potential of $8\,V$. First, the current was gradually increased till $45\,A$ and maintained for $40\,s$. Here the trimmed tip glowed dark red, and adsorbed gas was removed. During degasing, the pressure rose to approx. $1 \cdot 10^{-4}\,mbar$. The degasing procedure was repeated two times.

For evaporation, the current was increased to $85 \ldots 95\,A$ until sparks emitted from the trimmed tip of the carbon rod. Usually, the tip broke after a few seconds of evaporation due to tension. Then the evaporation was stopped. Hence, it was only possible to obtain film-thicknesses of $3 \ldots 8\,nm$ (see 3.1.2) in one step. The distance between the rod-tip and the sample was usually kept at $7.5\,cm$.

In case of the $e^-$-beam evaporation, a MED010 coating system (BAL-TEC AG) equipped with a head for e$^-$-beam-evaporation was used. As evaporation-source, a carbon-rod (Ø3.05 $mm$ BAL-TIC GmbH, spectrally clean) was used. Here a high-voltage was applied between the carbon-rod and a tungsten-filament that surrounded the rod in a distance of approx. $2\,mm$. The end of the rod was positioned in the center of the three-fold coiled filament. At a pressure of $1 \cdot 10^{-5} \ldots 2 \cdot 10^{-5}\,mbar$ the rod was degased using a potential of $1500\,V$ at $50\,mA$ for $5\,min$. After a pressure of $8 \cdot 10^{-6}\,mbar$ was reached, the evaporation was performed at $1800\,V$, $85 \ldots 90\,mA$ and a distance of $19\,cm$ between sample and evaporation-source. In contrast to thermal evaporation, much thicker carbon-films could be produced with $e^-$-beam evaporation. For both, thermal- and $e^-$-beam-evaporation, the obtained carbon-film was amorphous, which was proven by electron-diffraction (see subsection 3.1.5). All parameters used for evaporation of carbon are listed in table 3.1.

| PVD-technique | deposition-parameters | | | source-specification |
|:---:|:---:|:---:|:---:|:---:|
| | pressure $[10^{-5}\,mbar]$ | current $[mA]$ | voltage $[V]$ | |
| thermal[a,b] | 1.0 | $85 \ldots 95$ | 8 | rod Ø6.15 $mm$, BALTIC, spectrally clean |
| $e^-$-beam[c,d] | 0.8 | $85 \ldots 90$ | 1800 | rod Ø3.05 $mm$, BALTIC, spectrally clean |

[a]degasing: $1.0 \ldots 5.0 \cdot 10^{-5}\,mbar$, $45\,A$, $8\,V$, $3 \ldots 40\,s$
[b]distance between substrate and source = $7.5\,cm$
[c]degasing: $1.0 \ldots 2.0 \cdot 10^{-5}\,mbar$, $50\,A$, $1500\,V$, $5\,min$
[d]distance between substrate and source = $19\,cm$

**Table 3.1:** Parameters for carbon-evaporation.

As substrate for the carbon-film, the inner side of a freshly cleaved mica-sheet (Grade "V2" $25 \cdot 25\,mm$, Plano GmbH) was used. It has a pristinely clean surface. Mica is a sheet-silicate and the freshly cleaved surfaces exhibit nearly perfect basal cleavage. The surface has a roughness below $1\,nm$. Before floating the carbon-films on water, every edge of the mica sheet was scratched with a scalpel. Otherwise the film sticks to the mica and would not float properly. To avoid contamination of the carbon-film, bi-distilled water was used. The Petri-dish was cleaned using a detergent. Then it was washed several times with distilled water and afterwards with bi-distilled water. For picking the floated film, a TEM-grid was positioned underneath the water-surface directly below a carbon-film. Subsequently the grid was pulled out of the water thereby picking up a piece of the carbon-film. This is shown schematically in figure 3.1.

a.)    b.)

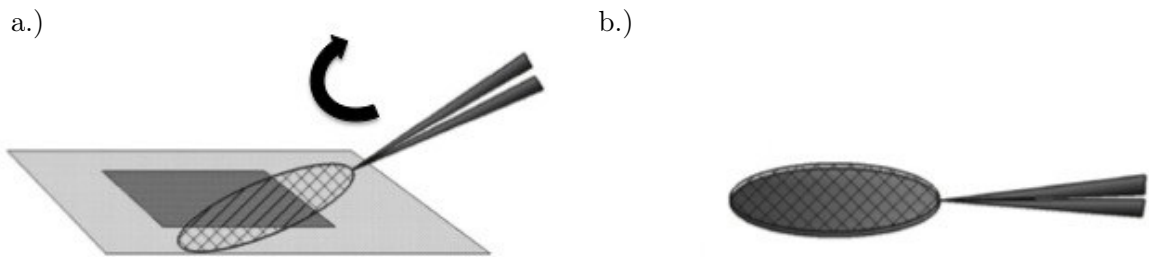

**Figure 3.1:** a.) A carbon-film is lifted from the water-surface with a TEM-grid; b.) TEM-grid with carbon film.

The grid was then left to dry on a slightly wetted filter paper for at least two days in an dessicator. For the first experiments, 200 mesh[1] copper-grids (Plano GmbH, ATHENE old) were used. Later experiments were made using gilded TEM-grids. This avoided contamination of the film, due to slight corrosion of the grid. Figure 3.2 shows two light-microscope images of carbon-films on different TEM-grids. In panel a.), a 200 square-mesh copper-grid, while in panel b.) a 100 hexagonal-mesh gilded grid was used as support for the carbon-film. Because the hexagonal shape is closer to the round shape of an aperture, hexagonal grids were taken for most experiments. Compared to commercial carbon support-films, e.g. from Quantifoil GmbH, the produced films are rather wrinkled. However, the production of very smooth commercial films usually involves the use of polymer-films and organic solvents that contaminate the carbon-film. Such organic contaminations cause charging in the TEM which renders a phase-plate useless. Several variations in the preparation of carbon-films were tried: a platinum wire-loop was used for lifting the film and putting it onto a TEM-grid; different TEM-grids were used and dry filter-paper was used for the drying of the films. However, the wrinkling of the films could not be avoided. Nevertheless, there were still a lot of windows of the TEM-grid, where the film was relatively plain and without fractures. Films in those windows could be taken for phase-plate preparation.

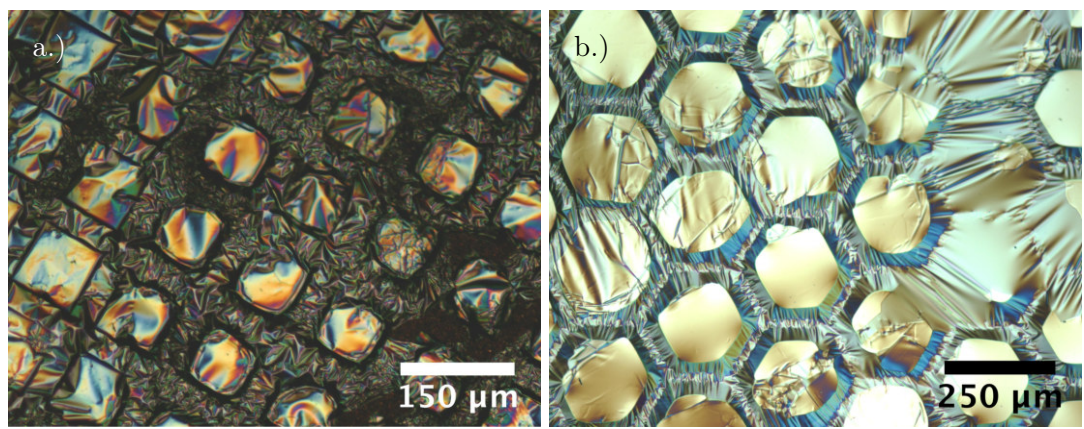

**Figure 3.2:** Images of carbon-films, taken with a light-microscope using differential interference-contrast (DIC): a.) Carbon-film on a 200 square-mesh copper-grid; b.) Carbon-film on a 100 hexagonal-mesh gilded copper-grid.

---

[1]Number of openings per inch (5.1 $mm$) of a mesh. The bar width of the TEM-grid has to be taken into account to calculate the size of windows.

### 3.1.2 Thickness-Measurement with Atomic-Force-Microscopy

The amount of phase-shift provided by a Zernike phase-plate, is proportional to its thickness. Therefore measuring the thickness of a thin-film was a major task during this thesis. Atomic-force-microscopy (AFM) was used to measure the thickness of both carbon- and metal-films. Hence thickness-measurement for both types of thin-film will be presented here.

AFM uses repellent Pauli-interaction (contact-mode) or attractive van-der-Waals-interaction (non-contact-mode) to make a topographic picture of a surface. For this, a nanometer-sized tip attached to a flexible spring-hanger (cantilever), is scanned over a surface using piezomotors. Through interaction of the nanometer-sized tip with the surface, the cantilever is bent and a laser-beam, pointed at a mirror on the rear side of the cantilever, is deflected. This deflection is measured by a four-zone photodiode to attain a value for the height as a function of the lateral position. Those informations are used to assemble a topographic image of the scanned surface. Figure 3.3 shows a schematic overview of the AFM.

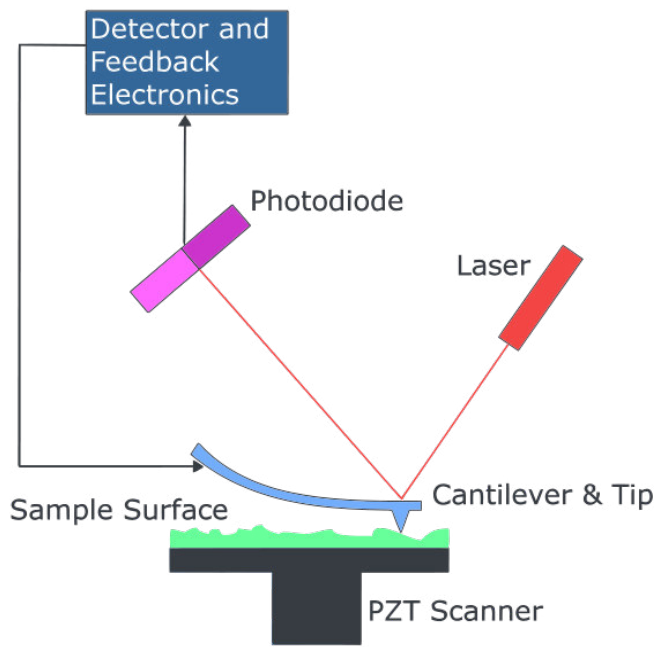

**Figure 3.3:** Schematic set-up of an atomic-force microscope (AFM)

All measurements were done using a VEECO dimension 3100 atomic-force-microscope operated in tapping-mode. Here the cantilever oscillates with a constant frequency near its eigenfrequency. While the tip is scanned over the surface, it taps on it and the amplitude and phase of the oscillation is changed. Those changes can be recorded and transformed into a height-signal. This signal is usually taken as a feedback for readjusting the height of the scanning tip during the measurement using a piezo-motor. This is done to avoid collisions of the tip with the surface, e.g. at a big step-edge, which would destroy the tip. The tapping-mode is commonly used for measuring under atmospheric conditions. AFM is a very accurate method to measure film-thicknesses, because of its high sensitivity for differences in height, as has been shown by Lemoine et al. They compared thickness-measurements of thin diamond-like films, performed with techniques like: AFM, visible-light interferometry, X-ray reflection, etc. They concluded that AFM while it is rather elaborate, is the most accurate method [33].

Measuring a film-thickness by AFM with $nm$ accuracy, requires a sharp step-edge between film and substrate. The step-width should be below $1\,\mu m$. If the step becomes too wide (several $\mu m$), the acquisition of data with AFM becomes problematic, and the step-height can be distorted by the slope of the underground.

For the generation of step-edges, several methods were tried: On a thoroughly cleaned glass-slide, a thin scotch-tape was attached. This was cut with a scissor on one side. After the whole sample was coated with thin-film, the scotch-tape was pulled off the glass-slide, producing the needed edge. The thickness of the film was measured with the AFM across this edge. For metal-films, step-edges were blurred using this technique and couldn't be measured with the AFM. Alternatively the step-edge was prepared by using a thin TEM-grid as covering object during deposition to create a step-edge. Also different substrates like polished silicon or a freshly cleaved mica-sheet were tried, but did not yielded better results. Figure 3.4 shows a blurred step-edge of an Au/Pd-film produced by covering a mica substrate with a thin 2000 mesh TEM-grid during evaporation.

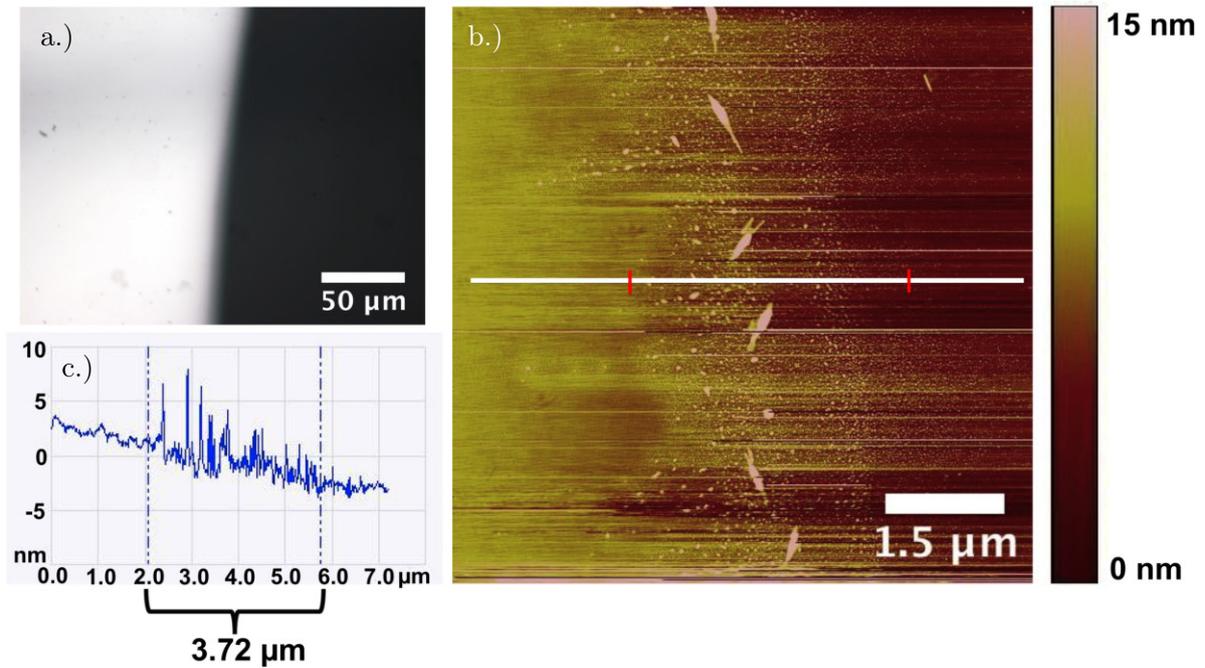

**Figure 3.4:** Step-edge produced by covering a mica substrate with a thin TEM-grid during evaporation: a.) Light-microscope image showing that the edge is blurred. b.) Topographic image of the same step-edge showing the broadness and roughness of the edge-area. The white line represents a line-scan height-profile shown in panel c.). Red lines mark the limits of the edge-area that was chosen as the area where the roughness of the sample is much higher than on the rest of the sample. Over this area of 3.71 $\mu m$ width, the information of the film-height gets lost.

Panel a.) shows a light-microscope image of the obtained step-edge. In panel b.) an AFM-image across the step-edge is shown. The AFM-image shows, that the edge is not confined, but rather consists of an area with a high roughness. Panel c.) shows a line-scan along the white line in the AFM-image. Here, the information about the step-edge gets lost in the broad and rough edge area that is approx. 3.72 $\mu m$ wide.

For measuring the thickness of metal-films, a second method was developed. Here, a thin mica-foil was pulled from a mica-sheet with scotch-tape, and attached to a freshly cleaved mica-sheet. It was important, that the edges of the mica-foil were not cut with a scissor, but ripped, so that the edges were fringy (see figure 3.6). Through van-der-Waal interaction, the mica-foil stuck very closely to the mica-sheet. After the film was deposited onto this substrate, the foil was gently pulled off using tweezers and the film-thickness was measured across a step-edge, produced by this procedure. Figure 3.5 gives an overview of the necessary steps for preparing mica samples for AFM thickness-measurements, hitherto called "sample for height-measuring"

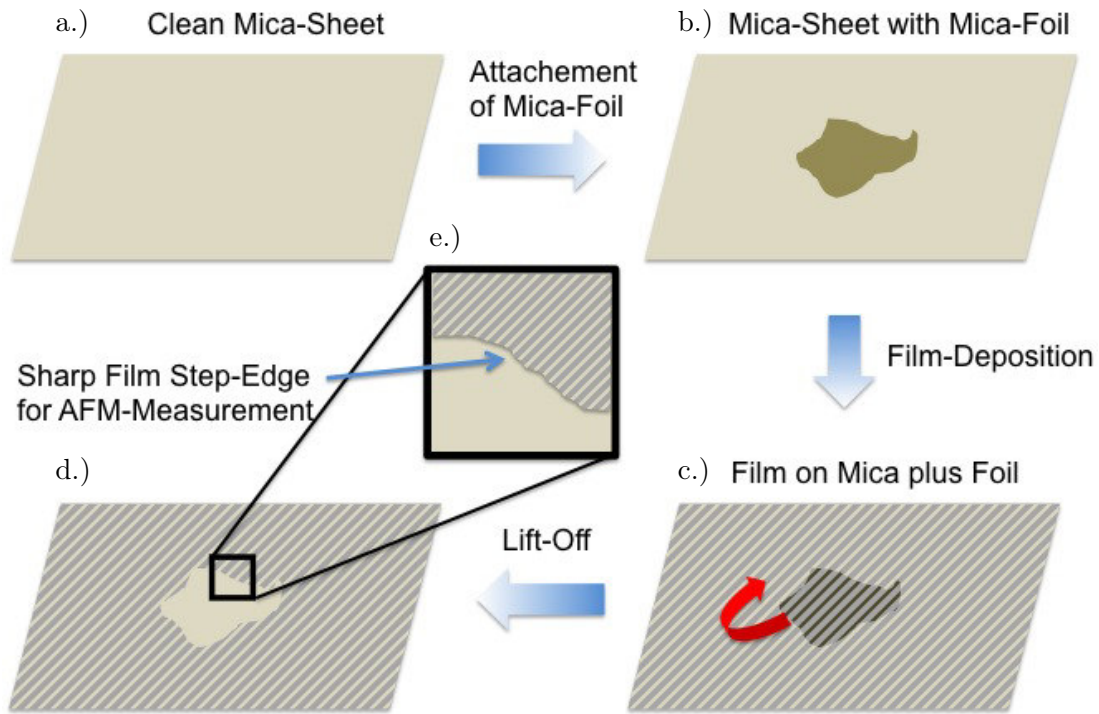

**Figure 3.5:** Necessary steps for preparing a mica sample for height-measuring ("mica-technique"): On a freshly cleaved mica-sheet (a.) a thin mica-foil is attached (b.). After a carbon- or a metal-film was deposited on the sample (c.), the mica-foil is pulled from the surface, leaving sharp film-edges on the substrate (d.). e.) is a magnified cut-out of d.) and shows the film step-edge where height-measurements are carried out with the AFM.

Light-microscope images of a step-edge produced with the "mica-technique" before and after removal of the mica-foil are shown are shown in figure 3.6.

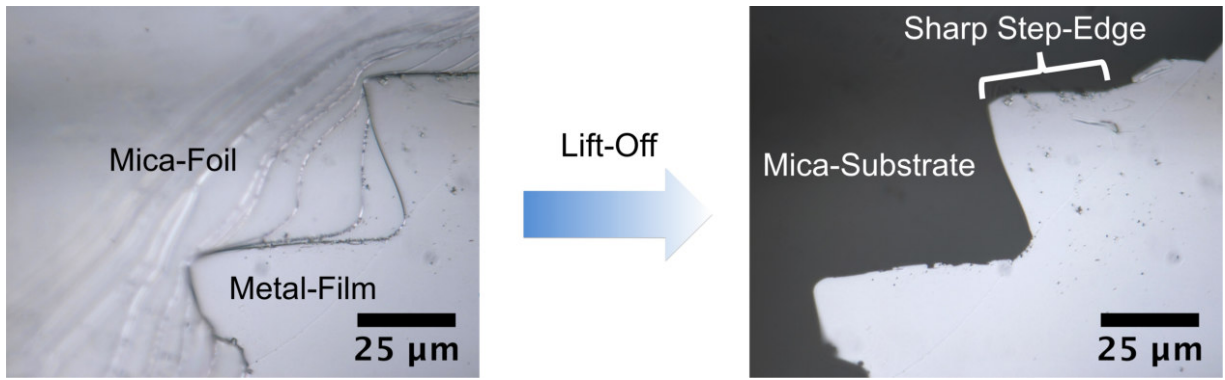

**Figure 3.6:** Light-microscope images of a mica sample for height-measuring before and after removal of a thin mica-foil. The two preparations steps are depicted schematically in figure 3.5 c.) and d.).

The produced step-edge is much sharper. This is also visible in the AFM-measurement across this step-edge, shown in figure 3.7 b.).

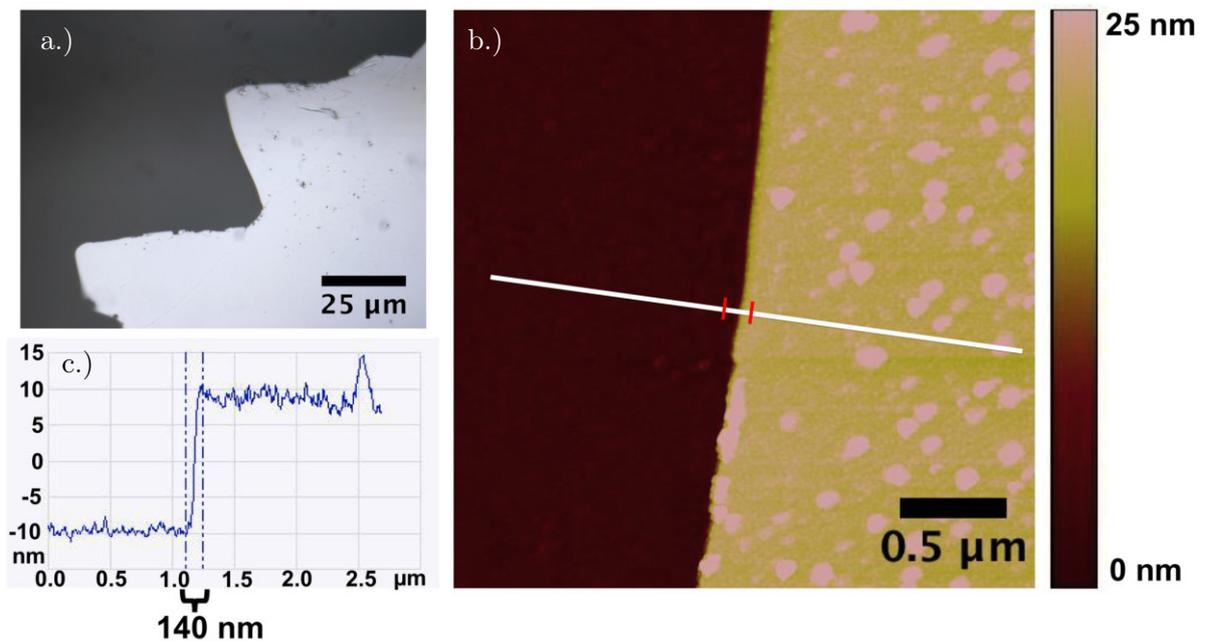

**Figure 3.7:** Images of a sharp step-edge produced by a mica-foil on a mica-substrate (see figure 3.5): a.) Light-microscope image of a sharp Al-film step-edge; b.) Topographic image of the step-edge shwon in a.). The white line marks the position of a line-scan height-profile shown in c.). Red lines mark the width of the step-edge, which is approx. $140\,nm$.

Here it becomes obvious, that step-edges produced with the "mica-technique" are considerably narrower compared to those produced with a scotch-tape or a thin object, covering the substrate during evaporation. The step-edge is approx. $140\,nm$ wide, according to the line-scan heigth-profile shown in panel c.). This is most probable due to the fact, that the mica-foil consists of many layers and that the rim of the foil is fringy as can be seen in figure 3.6 (left panel). The step-edge is formed by the undermost, nanometer thick layer of the fringy mica-foil, rather than by a micrometer-thick edge of a TEM-grid or a scotch-tape. Additionally, the mica-foil sticks to the mica-substrate very close which is also beneficial for a sharp step-edge.

Prior to thickness-measurements, the AFM was calibrated with a standard sample for height-measuring with a nominal step-height of $100\,nm$ (TED PELLA Inc.). A silicon $TESPA^{©}$-tip (Bruker Nanoscience) was used for all measurements. Film-heights were measured, by taking topographic images of the edge-region with a scan-size of usually $3 \cdot 3\,\mu m$. Before extracting the step-height, the slope of the sample was corrected, which would distort the measured film-height. For this, an area of the image that was assumed to be flat was selected. Here the slope of the sample was obtained. Afterwards, this slope was substracted from all data-points, yielding a slope-corrected topographic-image. The film-height was then obtained, by using the "rotating-box" function of the microscopes software. With this, a bigger region of the topographic-image containing the step-edge, could be averaged rotationally, yielding the step-height averaged over this region. Figure 3.8 shows the same topographic image shown in 3.7.

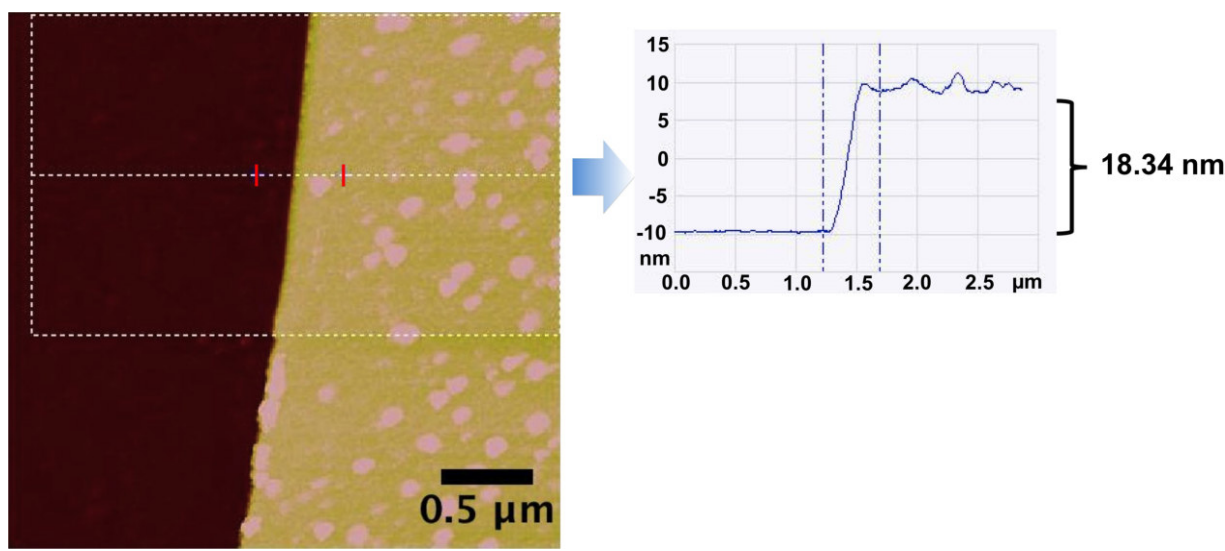

**Figure 3.8:** Topographic image of a step-edge of an Al-film. The "rotating-box" is represented by the dashed rectangle with a dashed central line. The red lines mark positions, between which the height-difference was calculated. The corresponding averaged height-profile is on the right. The film-thickness was determined to $18.34\,nm$.

The dashed-white rectangle represents the "rotating-box". A large number of crystallites near the edge could distort the height-information. Accordingly, an area with fewer crystallites was chosen to carry out the calculation. Positions between which the height was measured, are represented as red and as dashed lines in the topographic image and in the height-profile respectively. They were chosen such, that the film-height represents the height difference between the substrate and the film-plane, without contributions by small crystallites (see height-profile on the right). Results of AFM thickness-measurements for carbon and metal-films are summarized in table 3.2.

| material | nominal-thickness[a] $[nm]$ | measured-thickness $[nm]$ | average-thickness $[nm]$ | standard-deviation $[nm]$ |
|---|---|---|---|---|
| C | - | 19.18, 20.97, 20.65 | 20.27 | $\pm 0.779$ (3.8%) |
| Al | 17.00 | 18.37, 17.96, 17.53 | 17.95 | $\pm 0.343$ (1.9%) |
| Au/Pd (80/20%) | 4.8 | 15.82, 14.62, 14.44, 14.65 | 14.88 | $\pm 0.547$ (3.7%) |

[a]measured with a quartz-crystal thickness-monitor

**Table 3.2:** Film-thicknesses measured with AFM.

Different thickness-values for one film were obtained at different positions lying $0.5 \ldots 1\,cm$ apart on the sample for height-measuring.

Thickness-values measured with the thickness-monitor and with AFM, were often in no good agreement. This can be seen in table 3.2 in case of the Au/Pd-film. During deposition, the sample for height-measuring was always positioned near the substrate and the quartz-crystal thickness-monitor, to ensure that deposition-conditions on each object were as similar as possible. An overview of the arrangement before vapor-deposition is shown in figure 3.9.

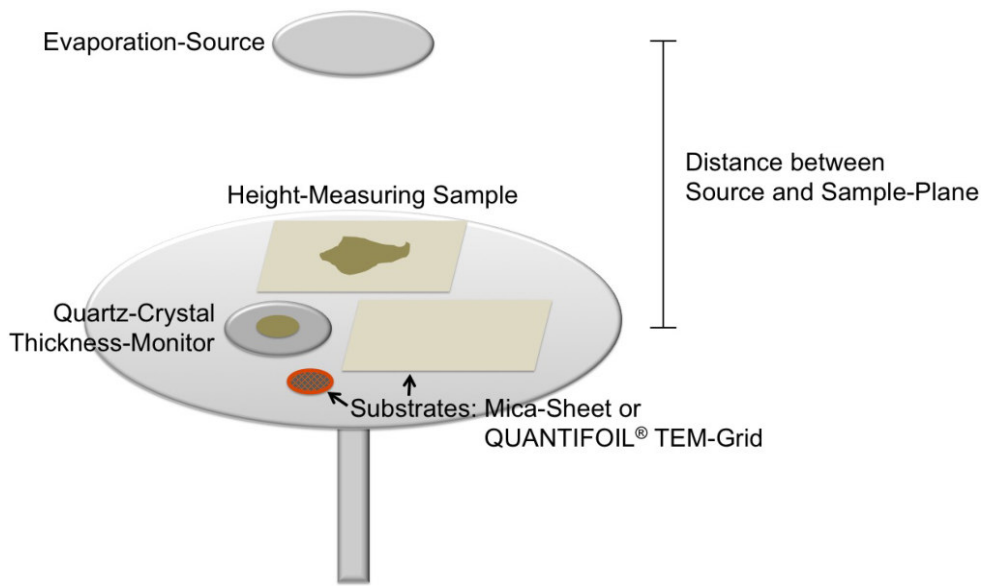

**Figure 3.9:** Overview of the experimental set-up before a deposition-experiment.

The distance between each object in the sample-plane and the evaporation-source is slightly different. The difference becomes bigger with decreasing distance between source and sample-plane. However, it is unlikely that the different deposition-conditions between sample for height-measuring and thickness-monitor account for the disagreement between thickness-values obtained with thickness-monitor and AFM. This is because positions on the sample for height-measuring where AFM-measurements were carried out, were approx. $0.5 \ldots 1\,cm$ apart. Those distances are in the same dimension than the distance between sample for height-measuring and thickness-monitor. The devitation is rather due to the inaccuracy of the thickness-monitor for the used deposition set-up. A general problem of quartz-crystal microbalances is the load of material already deposited on it. For increasing loads, the microbalance becomes less sensitive for further depositions, because the change of the resonance-frequency decreases. If the thickness-monitor is positioned at the bottom - as was done here - particles e.g. dust accumulate ontop of the quartz-crystal and distort the measurement. Additionally, high deposition-rates achieved with sputtering are problematic, because the load of the quartz-crystal is changed quickly.

A better setup was used in case of the Al-film listed in table 3.2. Here, the distance between source and substrate was large ($70\,cm$) and the source was at the bottom, while thickness-monitor and substrates hang upside down on the top of the evacuated chamber. The nominal-value obtained with the thickness-monitor is in good agreement with the value measured with AFM in this case. For setup used here, film-thicknesses should always be controlled by choosing the right deposition parameters and time, at least for films with thicknesses below $100\,nm$. To avoid a distortion due to a high load already deposited on the quartz-crystal, it is advisable to take fresh quartz-crystals.

For carbon-films, step-edges produced with the "mica-technique" and by usage of a scotch-tape were compared. Both techniques yielded sharp edges, but heights of step-edges produced with a scotch-tape were approx. twice as big than those, produced with the "mica-technique". It could be, that by using a relative thick scotch-tape, the film rolls up when the tape is removed, due to relative high forces used here, and is folded at the edges. Therefore AFM-measurements using mica sample for height-measurings, is best for all types of thin-films examined in this thesis.

### 3.1.3 Phase-Plate Preparation

The calculated thickness of a carbon-film for a phase-plate providing $-\pi/2$ phase-shift at $200\,kV$ is $23.71\,nm$. To reach this thickness, multiple depositions of carbon were needed using thermal evaporation, because with this only film-thicknesses ranging from $3 \ldots 8\,nm$ could be achieved at one time. The additional layers were deposited from both sides of the starting layer, to cover contaminations on the core-film from the floating-process. The thickness of each layer was measured with AFM yielding a total thickness of $29.8\,nm$. This should yield a phase-shift of approx. $-0.63\,\pi$.

The central hole was milled into the phase-plate film with a focused ion-beam (FIB). This was done at a Zeiss 1540XB focused ion-beam-system equipped with a scanning electron-microscope (SEM). It uses a focused beam of Ga$^+$-ions to cut structures into a sample. Figure 3.10 shows SEM-images of the phase-plate and the central hole.

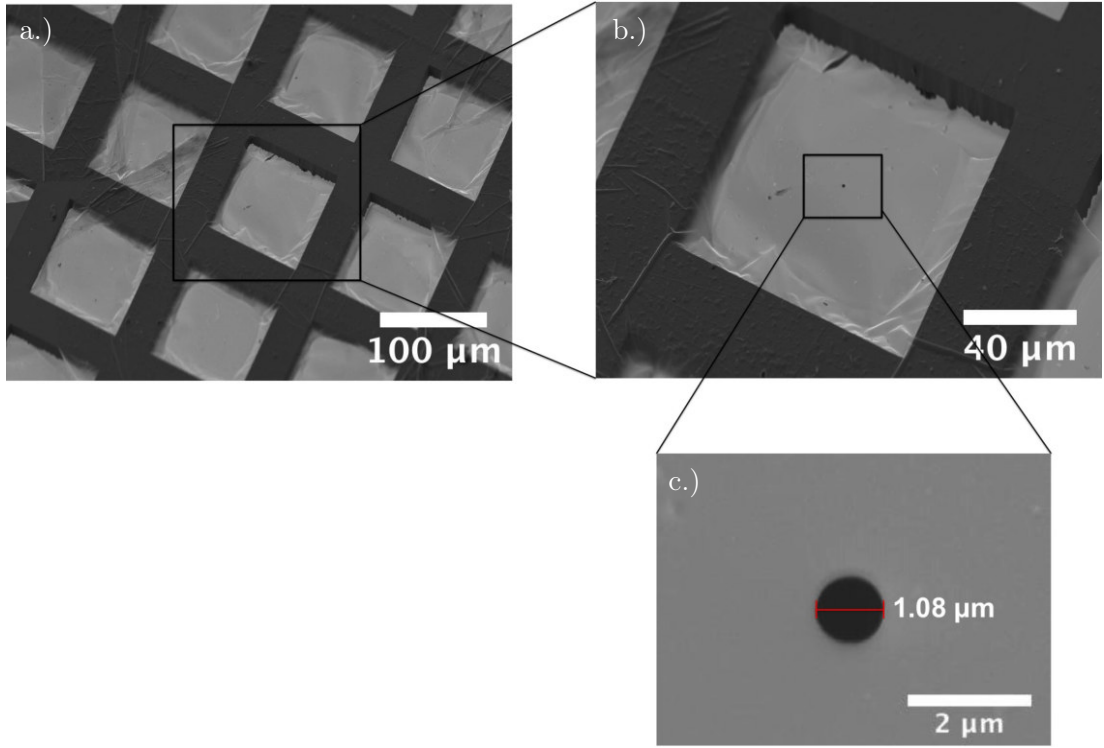

**Figure 3.10:** SEM-micrographs of a carbon phase-plate. a.) shows an overview of the TEM-grid with the carbon-film. Note that the film consists of several layers which can be seen at the rims of the windows. Panel b.) shows that the hole is positioned in the center of a TEM-window. The size of the central hole (c.) is $1.08\,\mu m$, and therefore slightly bigger than the nominal value of $1\,\mu m$.

Carbon phase-plates were made using both thermal and $e^-$-beam evaporated core-films. Also on the $e^-$-beam deposited carbon-film, one additional layer was deposited on each side (wrapping). Unfortunately, those phase-plates could not be tested, due to technical problems.

### 3.1.4 Application as Phase-Plate

For phase-plate tests a thin carbon-film was taken as sample. It can be considered a phase-object. For easier focusing, the sample had a cross-grating structure with gold-particles. The pictures were taken at a high defocus of about $-1000\,nm$, to obtain many thon-rings in the Fourier-transformation (FFT) yielding many extrema in the corresponding power-spectrum. Those ex-trema are needed to be able to fit the squared modulus of the $CTF$ properly to the profile of the FFT (power-spectrum). To insert the phase-plate into the back-focal-plane, a phase-contrast system of KonTEM GmbH was used. It consists of a rod, similar to a TEM specimen-holder and can be attached to the TEM-column instead of the objective-aperture holder. The phase-plate is held by a small tip that is attached to the end of the rod. By the use of piezo-motors, the phase-plate can be positioned in the optical axis with a nanometer accuracy.

The alignment procedure was done as follows: By the use of the combined spot-diffraction-mode of the microscope, the phase-plate, lying near the back-focal-plane of the objective-lens, could be imaged and was roughly centered in the beam-axis. In the normal image-mode, the central hole appeared as a bright spot, projected onto the sample. By changing the current of the condensor-lens "C3", this bright spot was magnified till it occupied the whole image. In doing so, the back-focal-plane was put precisely to the same height, as the phase-plate. At the same time, the central hole was aligned with the beam-axis, by moving the phase-plate with the phase-plate holder and hence keeping the bright spot in the center of the image. A ZEISS Libra 200 "corrected illumination scanning probe TEM" operated at $200\,kV$ was used for all phase-contrast experiments.

To quantify the phase shift induced by a phase-plate, two images of a phase-object, have to be recorded at the same defocus-value. One with, and the other one without phase-contrast. Afterwards, both images are Fourier-transformed. The profile of each FFT yields a power-spectrum, that is equal to the squared modulus of the $CTF$ ($\left|CTF^2\right|$). By fitting $\left|CTF^2\right|$ to the power-spectrum of the "no-phase-contrast" image, the exact defocus can be determined. This defocus is taken as input for fitting the squared modulus of the phase-contrast $CTF$ (see equation 3.3) to the "phase-contrast" power-spectrum. In doing so, the phase-shift induced by the phase-plate can be obtained.

Danev et al. tried to determine the phase-shift induced by the phase-plate by comparing TEM-micrographs for the aligned and for the retracted phase-plate. With this procedure, no phase-contrast phase-shift could be obtained. This procedure is inadequate for determination of the phase-contrast phase-shift, because of influences of the thin-film. When the continuous thin-film of the phase-plate is positioned in the back-focal-plane, a large shift of the power-spectrum is observed, even though no phase-contrast is induced. This is shown in figure 3.11. Although there is no, or just a very slight change visible in the TEM-micrographs when inserting the phase-plate-film, the difference is obvious in the corresponding FFTs. Power-spectra shown in the undermost panel, were obtained by averaging the profile of the FFTs over a rotation-angle of $90°$ in the first quadrant, with the TIA imaging software for electron-microscopy. Profiles were background-subtracted using $13\ldots17$ points of support, set manually and normalized to unity. The red arrow, illustrates the shift when inserting the thin-film.

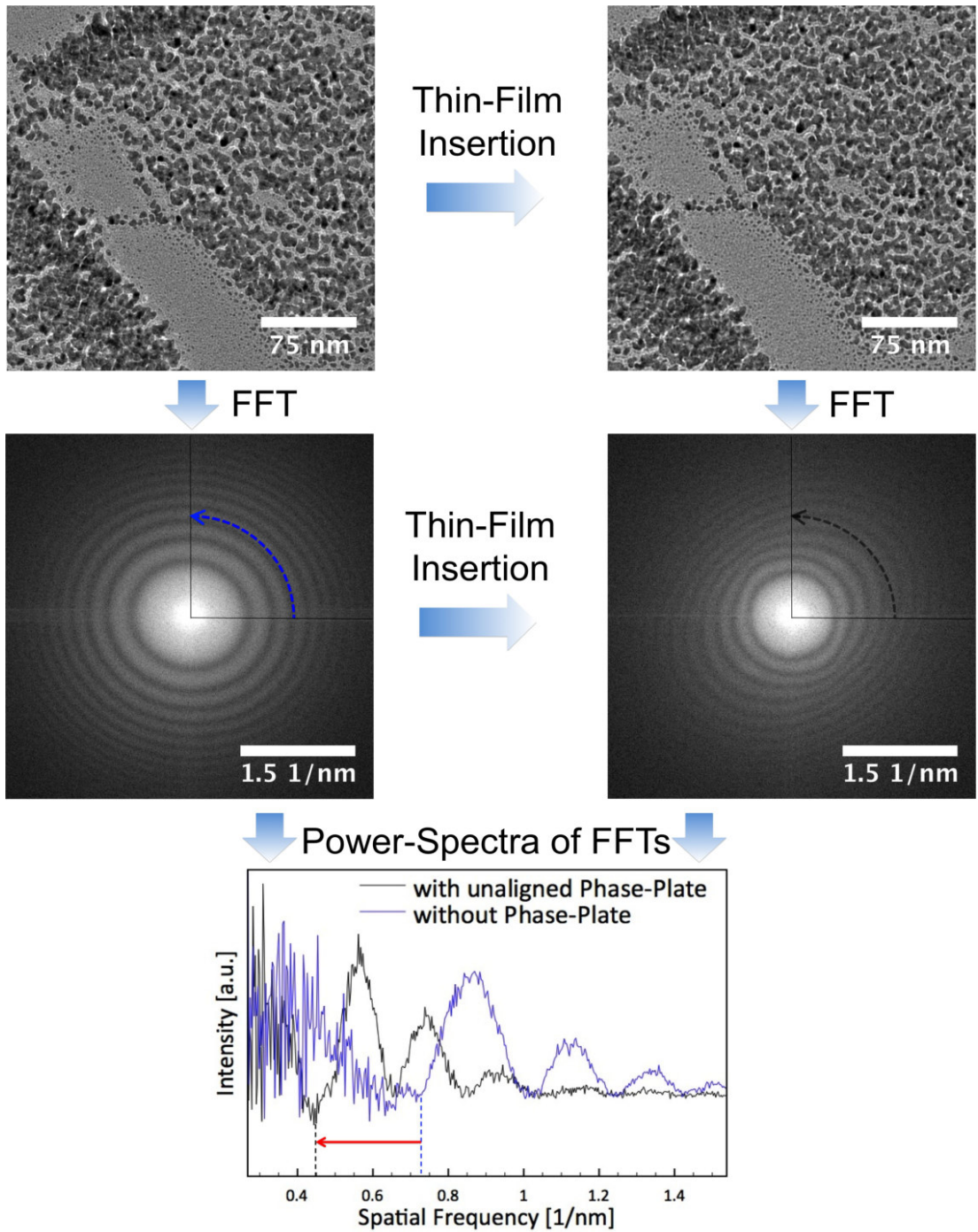

**Figure 3.11:** TEM-micrographs and corresponding FFTs of a reference-sample with the same high underfocus, without (left) and with unaligned phase-plate in the back-focal-plane (right). A change in the corresponding FFTs similar to a big increase of underfocus is observed. Bent arrows represent the regions and the angle (90°) over which the profiles were averaged, yielding the power-spectra given in the diagram below. Dashed lines mark the positions of the first minima, which are clearly shifted as indicated by the red arrow.

The shift in the power-spectrum is similar to an increase of underfocus. This could be explained as a lens-effect of the thin-film, induced by charging. Such an effect was reported also for the Boersch phase-plate by Majorovits et al.[10]. Here, the miniature einzel lens induces a phase-shift proportional to $k^2$. This phase-shift can be subsummed under the defocus-term of the $CTF$ that is also proportional to $k^2$, and thus leads to a pseudo-change of defocus. Charging of carbon-films in the back-focal-plane was also reported by Danev [17, 18]. Danev et al. found out, that the charging is positive and decreases radially from the position of the beam-spot [34]. To verify, if the phase-shift induced by the thin-film could be explained by a pseudo-change of defocus, $\left|CTF^2\right|$ was fitted to the obtained power-spectra. This is shown in figure 3.12. Here power-spectra for the case of the retracted and the inserted, unaligned phase-plate are shown. The intensity-scale is logarithmic so that minima are shown more clearly. For clarity, only the minima-positions of the calculated $\left|CTF^2\right|$ are shown, represented as dashed lines. In case of the retracted phase-plate, $\left|CTF^2\right|$ could be fitted fairly to the power-spectrum (left). The spherical-aberration was $1.2\,mm$ and the electrons had an energy of $200\,keV$. The defocus obtained by fitting is $-780\,nm$. The same could not be done satisfactorily for the inserted thin-film, when a change of defocus was included. From the fourth minimum (marked by red arrow) onwards, the minima do not coincide. A defocus of $-1800\,nm$ was used for the calculation of the minima. While the phase-plate was centered in the beam, the FFT changed rapidly with time, which was in agreement with reports of Danev et al., who assigned this change to a charging of the phase-plate that reaches a quasi equilibrium-state after $30\,min$ [17].

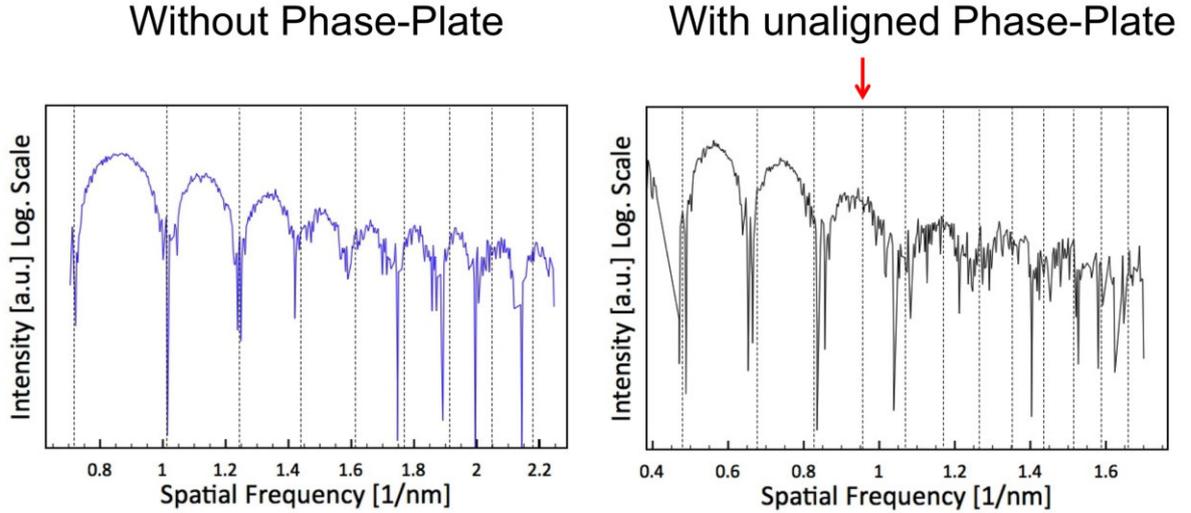

**Figure 3.12:** Power-spectra of TEM-micrographs shown in figure 3.11 without (left) and with unaligned phase-plate in the back-focal-plane (right). Note that the intensity-scale is logarithmic so that minima are more pronounced. The minima of a calculated $\left|CTF^2\right|$ (dashed lines) can be fitted fairly to the left spectrum, with spherical-aberration of $1.2\,mm$, electron-energy of $200\,keV$ and defocus of $-780\,nm$. In case of the right spectrum, from the fourth minimum (marked by red arrow) onwards, the fit is not satisfactorily.

To account for the distortion by the thin-film, the phase-contrast phase-shift was tried to determine, by comparing TEM-micrographs with the aligned (phase-contrast) and unaligned (no phase-contrast) phase-plate. Results are illustrated in figure 3.13. Images and spectra were obtained and processed equally to the previous ones. There is a remarkable change of contrast visible in the TEM-micrographs. FFTs are compared directly in panel c.). The red arrows mark the displacement of the thon-rings indicating a phase-shift. This is also visible when observing the corresponding power-spectra. Curiously, a positive phase-shift is applied, while it should be a negative.

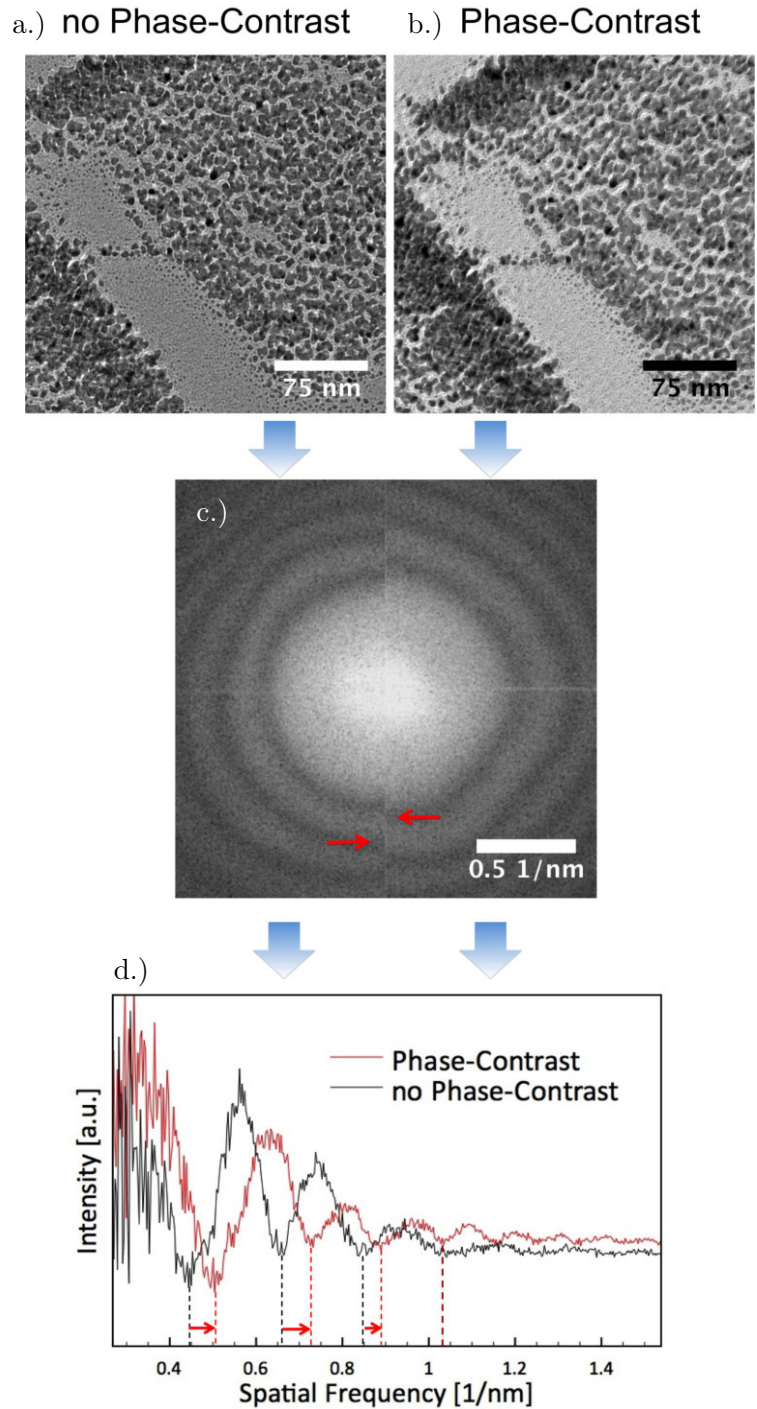

**Figure 3.13:** Bright field TEM micrographs of a cross grating sample. a): Phase-plate in the back-focal-plane without aligning the central hole. b): Phase-plate in the back-focal-plane, central hole aligned. Both images were taken with the same settings for defocus. c.): Superimposed FFTs of images A (left) and B (right). Red arrows mark the displacement of the thon-rings. d.) shows scans of the FFTs averaged over 90° of rotation. Dashed-lines mark minima of the power-spectra, which are clearly apart from each other. Curiously the phase-shift applied has the wrong sign.

One explanation for the positive phase-shift might be, that an additional minimum for the phase-contrast FFT is covered by the unscattered beam (bright central spot in the FFT). Another reason might be, that charging for the unaligned phase-plate is much higher than for the aligned. In this case, the phase-shift due to the charging could overcompensate the phase-contrast phase-shift.

Although the phase-shift induced by the phase-plate has the wrong sign, the phase-contrast effect of the phase-plate is still visible (see panel a.) and b.). Further experiments have to be carried out, to be able to provide a definite reason for the positive phase-shift induced by the phase-plate.

### 3.1.5 Characterization with TEM

TEM-micrographs as well as electron-diffraction-pattern were used to analyze the microstructure of the produced carbon films. Electron-diffraction can be used to determine the crystalline structure of a solid. Electrons used in conventional TEM with an energy of $100\ldots300\,keV$, have wavelength ranging from $3.7\ldots2.0\,pm$. Atomic distances in a solid are in the same range. Therefore, crystal-lattices can be used as a diffraction-lattice for those electrons.

When electron-waves are scattered by atoms of a crystal, they undergo constructive and destructive interference. Only if the distance between lattice-planes by which electron-waves are scattered, is an integer multiple of their wavelength, the electron-waves are not erased by destructive interference. For those lattice-planes where constructive interference occurs, electron-waves are reflected. For a crystalline sample, this leads to sharp diffraction-spots in the diffractogram. In a polycrystalline sample, concentric diffraction rings occur instead of separated spots, due to the statistically distributed orientation of all lattice-planes. Sharp diffraction-rings or spots correlate to small angular-distributions of the corresponding lattice-planes. The less crystalline a solid is, the broader is the distribution for the scattering-angle of one specific lattice-plane and the broader and less intense is the corresponding diffraction-reflex. Hence an amorphous solid, leads to very broad diffraction-rings, with low intensity.

In TEM, electron-scattering is usually done with a small diffraction-aperture, limiting the area of the sample that is used for the diffraction-experiment. This area, can be chosen, so that diffraction is carried out on a selected object, for example a nano-crystallite. Therefore, the according technique is called selected-area-diffraction (SAD).

Figure 3.14 shows a TEM-micrograph and the corresponding SAD-image of a carbon phase-plate. Images were taken with a JEOL JEM 2200FS operated at $200\,kV$. The SAD-image was taken with a camera-length of $30\,cm$ and a small diffraction-aperture was used. The d-spacing was calibrated, using the distances of the Au(111) lattice-plane measured with a polycrystalline Au-sample as reference.

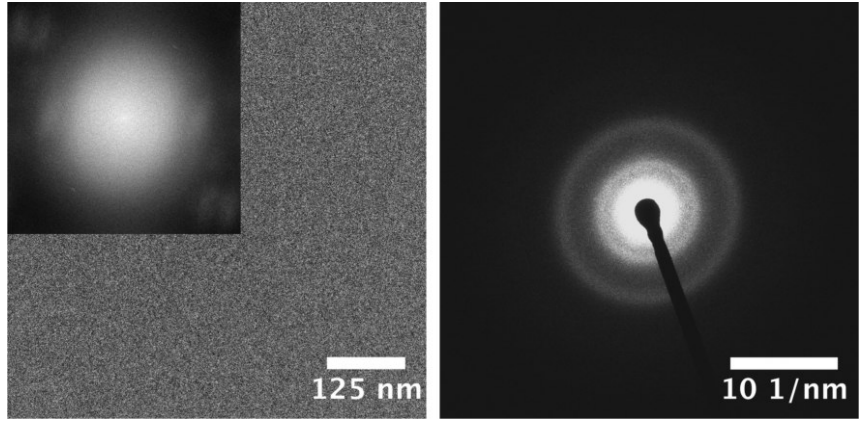

**Figure 3.14:** TEM-micrographs (left) and selected-area-diffraction images (right) of a carbon phase-plate. The inset shows the Fourier-transformation of the TEM-micrograph indicating that the corresponding image is taken near focus. Note that the carbon-film exhibits no crystalline structure. The SAD-image shows that the film is highly amorphous, because diffraction-rings are very weak and broad.

The phase-plate had an $e^-$-beam evaporated core and thermal-evaporated "wrapping". The TEM-micrograph shows no micro- or nano-structure. It was taken near Gaussian-focus, which is shown by the FFT (inset) exhibiting no thon-rings. According with the TEM-micrograph, the SAD-image reveals that the film is highly amorphous, showing only weak, broad diffraction-rings.

### 3.1.6 Discussion and Outlook

Carbon phase-plates have been prepared according to Danev et al. [18]. The thickness of the films was measured using AFM. Here, an improved technique for the preparation of sharp step-edges was developed. An advantage of this technique is that the film-thickness can be measured relatively quick, without examining the according film with a transmission-electron-microscope. A disadvantage of AFM-measurements is, that the film-thickness is always measured on a sample for height-measuring, whereas with electron energy-loss spectroscopy (EELS), the thickness is measured directly at the thin-film of interest. This possible uncertainty can be reduced, with a good set-up of the deposition-experiment, e.g. a long distance between source and sample-plane. It could be shown that inserting the thin-film of the phase-plate into the back-focal-plane, causes a change in the imaging-conditions. This change couldn't be accounted for by adding a phase-shift proportional to $k^2$ and thus by changing the defocus-value. Results of experiments suggest that when a thin-film is inserted into the back-focal-plane, the $CTF$ has to be modified:

$$CTF(k) = \sin[\gamma(k) + \phi_{dist,f}(k)] \tag{3.1}$$

Here, $\phi_{dist,f}(k)$ is a phase-shift $k$-dependent, caused by distortions of the continuous thin-film in the back-focal-plane. One possible source of those distortions is charging of the thin-film. Furthermore, $\phi_{dist}(k)$ can be probably assumed as a sum of a lens-effect term $\phi_{lens}(k)$ proportional to $k^2$, and another distortion-term $\phi'_{dist}(k)$ depending on $k$:

$$\phi_{dist}(k) = \phi_{lens}(k) + \phi'_{dist,f}(k) \tag{3.2}$$

with $\phi_{lens} \propto k^2$

When the phase-plate is aligned on the optical axis, the $CTF$ can be modified accordingly:

$$CTF(k) = \sin[\gamma(k) + \phi_{dist,c}(k) + \phi_{PC}] \tag{3.3}$$

Here $\phi_{dist,c}(k)$ is a phase-shift caused by distortions of the centered thin-film in the back-focal-plane and $\phi_{PC}$ is the constant phase-shift due to the phase-contrast effect of the phase-plate known from equation 3.3. Further experiments have to be carried out, to determine the distortion by the thin-film precisely.

Danev et al. suggested, that the charging of the carbon phase-plate is partly caused by non-conducting residues from the floating-process [18]. To avoid the floating-process and related contaminations, a process for the preparation of a carbon-film via plasma-cleaning was sketched. It was adopted from the process of metal-film preparation. The involved preparation-steps are shown schematically in figure 3.15.

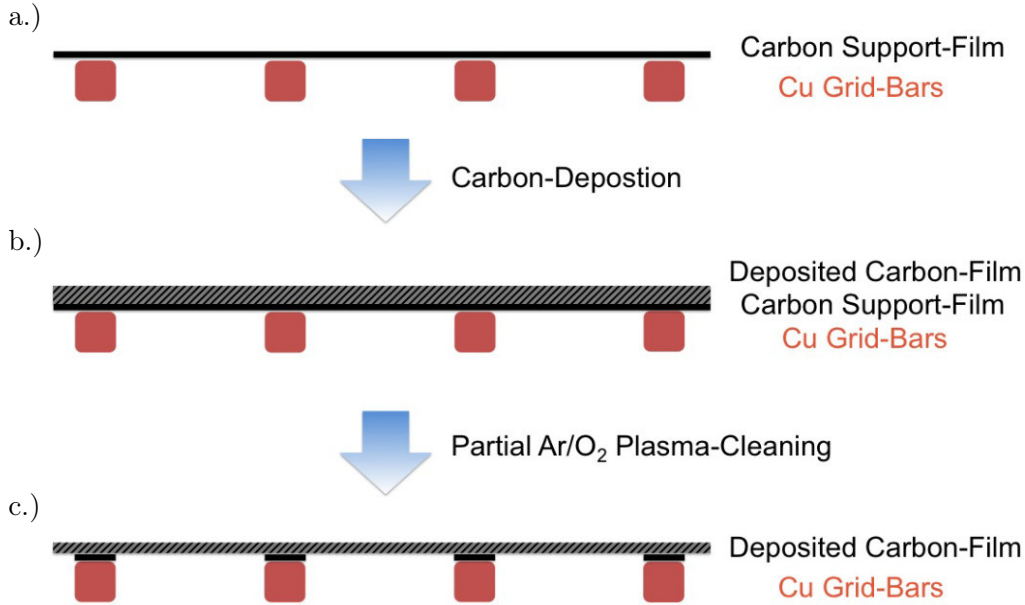

**Figure 3.15:** Overview of preparation-steps of an "improved" carbon-film without the usage of floating: Onto a commercial carbon support-film on a TEM-grid (a.) a carbon-film is deposited via $e^-$-beam evaporation (b.). Afterwards, the support-film and part of the deposited carbon-film is taken away by careful plasma-cleaning (c.).

Onto a commercial carbon support-film, another carbon-film is evaporated by $e^-$-beam evaporation. After the second carbon-film was deposited, the support-film can be taken away, by careful plasma-etching with an $Ar/O_2$-Plasma. The rate of the plasma-etching could be measured with thickness-measurements carried out with EELS. By etching away the support-film and additionally a certain part of the deposited carbon-film, it should be possible to remove all contaminations introduced by the support-film. The thickness of the deposited-film, could be chosen in such a way, that the remaining carbon-film has the right thickness for a $-\pi/2$ phase-shift.

## 3.2 Metal-Films

There are sound reasons why metal-films should be investigated as material for Zernike phase-plates. First of all, the higher electrical conductivity compared to carbon should largely eliminate charging. While graphite exhibits a high conductivity parallel to crystallographic-layers, the amorphous carbon films made during the experiments for this thesis have a specific-conductance ranging from $1.25\,S/m \ldots 2 \cdot 10^3\,S/m$ [35], and are poor electric conductors. Aluminium as a metal, has a specific conductance of $3.5 \cdot 10^7\,S/m$ [35]. The difference between the conductivity for those two materials is between $4 \ldots 7$ orders of magnitude.

Another reason for applying metal-films as Zernike phase-plates is, that beam-damage should be lower in comparison to carbon-films. Metals exhibit higher thermal conductivity than covalent materials like carbon, and are therefore less endangered by damage through heating of the sample [36]. In addition to this, carbon shows knock-on damage (displacement of atoms by electrons) above approx. $100\,kV$, while for Al it is far above $200\,kV$ [36].

Another important advantage of metals is, that they do not adsorb gas-particles on their surface, like in case of carbon. It was reported [37], that carbon phase-plates were heated in the TEM-column to degas them. This is not necessary in case of metal-films. In this section the preparation, characterization and application of metal-films as phase-plates will be presented.

### 3.2.1 Preparation

In contrast to carbon-films, metal-films can not be floated off from a substrate like mica, because they stick to it much stronger. There are other techniques to produce self-sustaining thin-films, but generally it is very difficult to produce such films out of metal.

The question was: how could a self-sustaining, thin metal-film be produced. The answer was given in a paper by Janbroers et al. [38] who developed a simple method to produce carbon-free sustaining-films for TEM. They took a conventional holey carbon-film on a TEM-grid and deposited an Au/Pd-film (80/20%) ontop of it, by sputtering. Subsequently, the carbon-support film was selectively taken away by plasma-cleaning in an $Ar/O_2$-plasma consisting of 75% Ar and 25% $O_2$. Such a plasma is commonly used, to clean samples for TEM of organic contaminations. He could show by using electron-energy-loss-spectroscopy (EELS), that with increasing time of plasma-cleaning, carbon was continuously removed, while the metal-film remained unaffected. All steps of metal-film preparation are shown schematically in figure 3.28. During all experiments, $QUANTIFOIL^{©}$ flat carbon 100 support films (Quantifoil Micro Tools GmbH) on 200 mesh copper-grids were used. Metal-films were deposited onto the support-film by either sputtering with Ar-ions, or by $e^-$-beam evaporation. Between grid-bar and metal-film, the carbon-film is most probably not removed completely, which should not affect the functionality of the phase-plate because the unsustained part is cleaned completely.

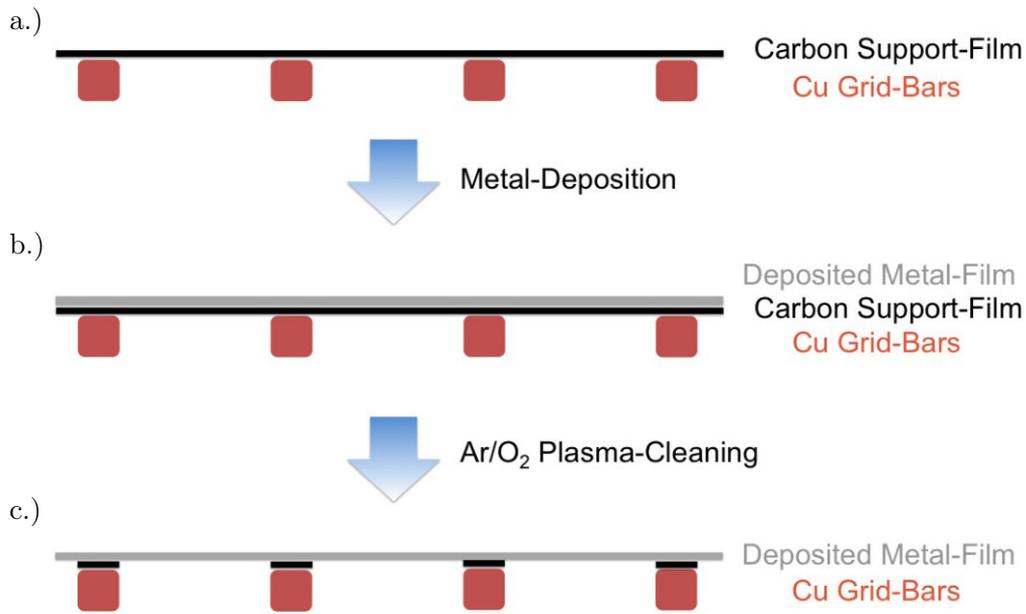

**Figure 3.16:** Overview of preparation-steps of a metal-film: Onto a commercial carbon support-film on a TEM-grid (a.) a metal-film is deposited via sputtering or $e^-$-beam evaporation (b.). Afterwards, the support-film is taken away selectively by plasma-cleaning (c.).

First experiments were carried out with an Au/Pd-alloy (80/20%) to reproduce the experiments by Janbroers et al. Later, Al and Cu were used as film materials.

Aluminium was chosen as material for phase-plate preparation, because of its relatively low Z-number, which reduces electron-scattering and thus loss of beam-intensity when used as a phase-plate. Due to less scattering, it even has a 10% higher electron-transmittance than carbon [17]. Furthermore, the mean inner-potential of aluminium is well known, in contrast to that of magnesium, which is the reason why aluminium was chosen instead of magnesium. Beryllium is the element with the lowest Z-number that is solid and can be handled at air. Because of its low absorbance for X-rays and electrons, it is used in many X-ray- and electron-microscopy devices. Furthermore, it provides good electric conductivity. All those aspects make beryllium the ideal material for Zernike phase-plates in theory. However, due to its toxicity and all difficulties related with it, beryllium was not chosen as phase-plate material for the work presented here. Thin copper-films were prepared in addition to the Al-films, because of the high electrical conductivity and relatively noble chemical behavior of copper. This could reduce charging in the TEM, because of a thinner oxide-layer. The parameters used for metal-film preparation are listed in table 3.3.

|  |  | sputtering-parameters | | |
|---|---|---|---|---|
|  |  | Ar-pressure | current[a] |  |
| material | PVD-technique | $[10^{-2}\,mbar]$ | [mA] | target |
| Al | sputtering[b] | 1.0 | 130 | Goodfellow, 99.0% |
| Al | $e^{-}$-beam[c] | - | - | Umicore, 99.999% |
| Cu | sputtering[b] | 1.0 | 60 | BALTIC, 99.999% |
| Au/Pd (80/20%) | sputtering[b] | 5.0 | 40 | BALTIC, 99.999% |

[a]high-tension = $250\ldots300\,V$
[b]distance between substrate and source = $7.5\,cm$
[c]distance between substrate and source = $70\,cm$; evaporation-rate: $1.2\,nm/s$; pressure: $3\cdot10^{-7}\,mbar$

**Table 3.3:** Parameters for metal-film preparation.

All sputtering-experiments were carried out using a Baltec MED020 (BAL-TEC AG) with a maximum high-tension of $1000\,V$. The distance between sputtering-target and sample was usually $7.5\,cm$. Al-films were also prepared by $e^{-}$-beam evaporation with a BAK 640 (Unaxis GmbH). Etching of the carbon-films after metal-deposition, was done in a model 1020 plasma-cleaner (Fischione Instruments). It uses an oscillating electromagnetic-field with a frequency of $13.56\,MHz$ to produce a plasma. The gas for plasma-cleaning was a mixture of 75% Argon and 25% $O_2$ (purity: 6.0 and 5.0, Linde AG). Metal-films were inserted into the plasma-cleaner, mounted on a TEM sample holder. During the process, the films were completely exposed to the plasma. The gas pressure during plasma-cleaning was kept between $4.9\cdot10^{-2}\,mbar$ and $5.6\cdot10^{-2}\,mbar$. After a period of 15 minutes, the carbon-film was completely removed, shown by EELS (see subsection 3.2.2).
Figure 3.17 shows light-microscope images of all preparation-steps of an Al-film, represented schematically in figure 3.16.

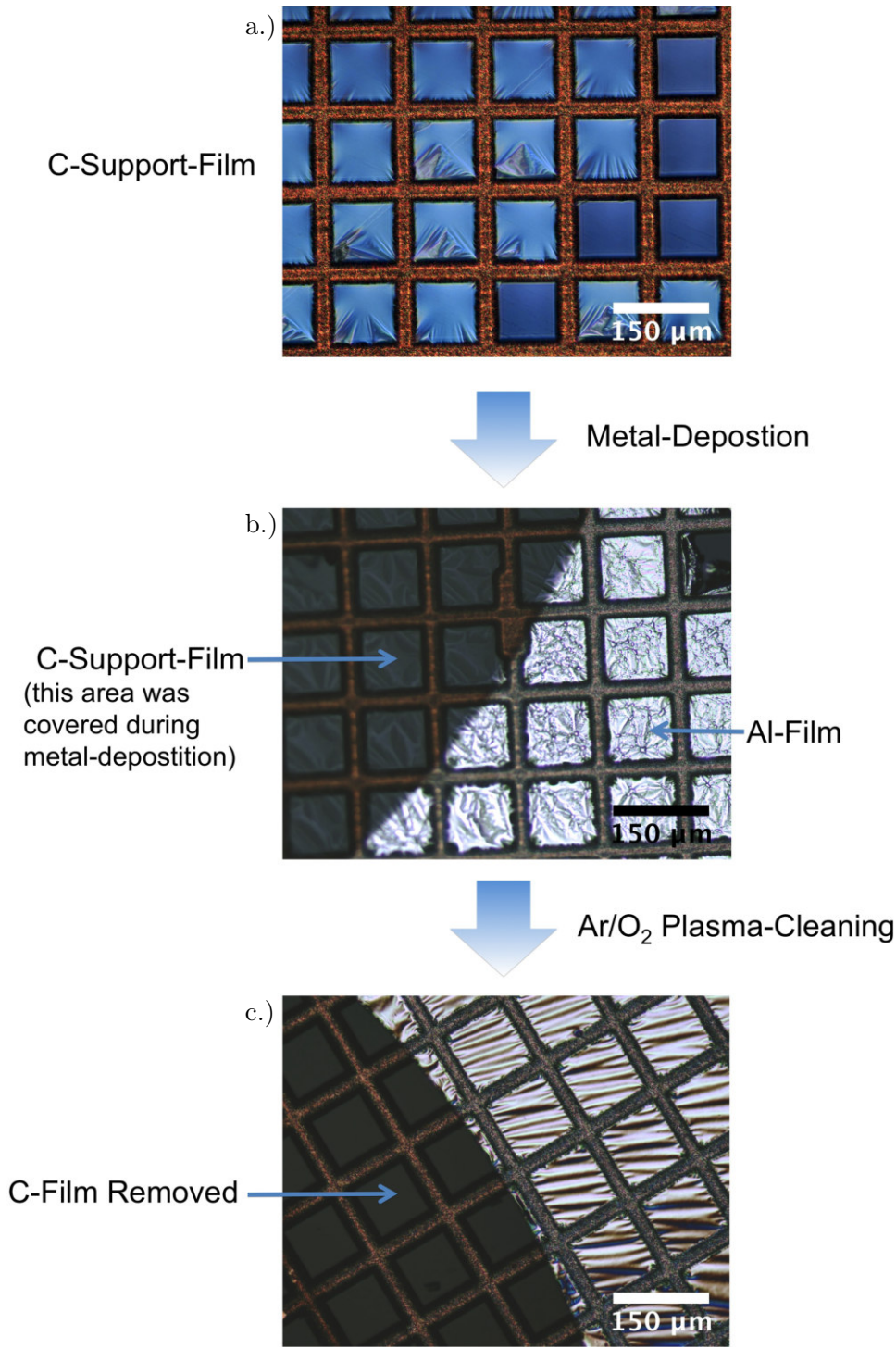

**Figure 3.17:** Light-microscope images (DIC) of preparation-steps shown schematically in figure 3.16. a.) shows a $QUANTIFOIL^{©}$ carbon support-film on a 200 mesh copper-grid. Note that this panel looks different than the others, because DIC was adjusted differently to make the carbon-film better visible. After metal-deposition the Al-film is visible on the right. Note that the area on the left side of panel b.) was covered during the deposition, so that the support-film is still visible. After plasma-cleaning (c.), there is no support-film in this area, showing that the carbon-film has been removed. While the Al- and carbon-film are wrinkled before plasma-cleaning, the Al-film appears wavy but smoother after it.

In panel b.) of figure 3.17, the support-film is still visible, because it was partly covered during metal-deposition. After the plasma-cleaning, the film is removed (panel c.). Additionally the support-film and the Al-film in b.) are strongly wrinkled, while after plasma-cleaning the Al-film is much smoother. This observation can be seen more clearly in figure 3.18. Here, the same Al-fim as shown in figure 3.17 is presented at higher magnification before and after plasma cleaning.

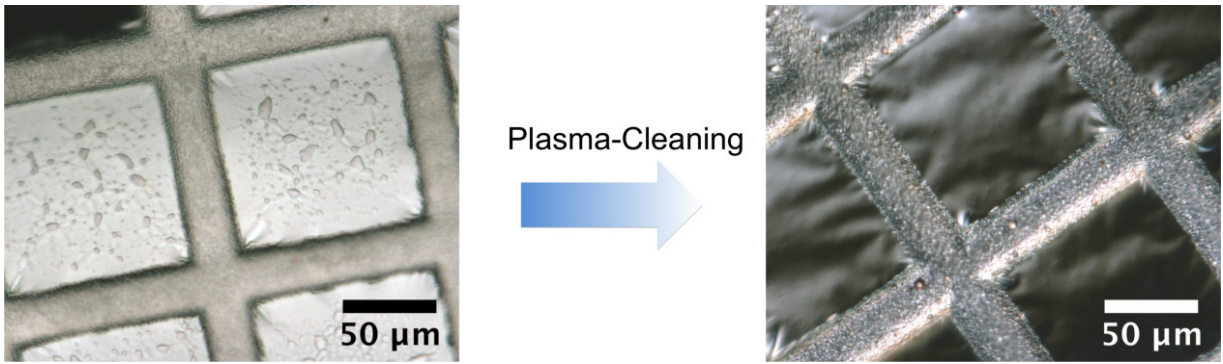

**Figure 3.18:** Light-microscope images (DIC) of an Al-film before and after plasma-cleaning. Before plasma-cleaning the film is wrinkled and afterwards smooth, indicating that wrinkles are caused by the support-film.

With other metal-films, this behavior could be observed as well. It is likely, that the support-film gets wrinkled during deposition due to the kinetic energy of the deposited particles, causing the formed-film to be wrinkled as well. After removing the support film, the metal-film relaxes and exhibits a smoother surface. Light-microscope images of all investigated metal films are shown in figure 3.19.

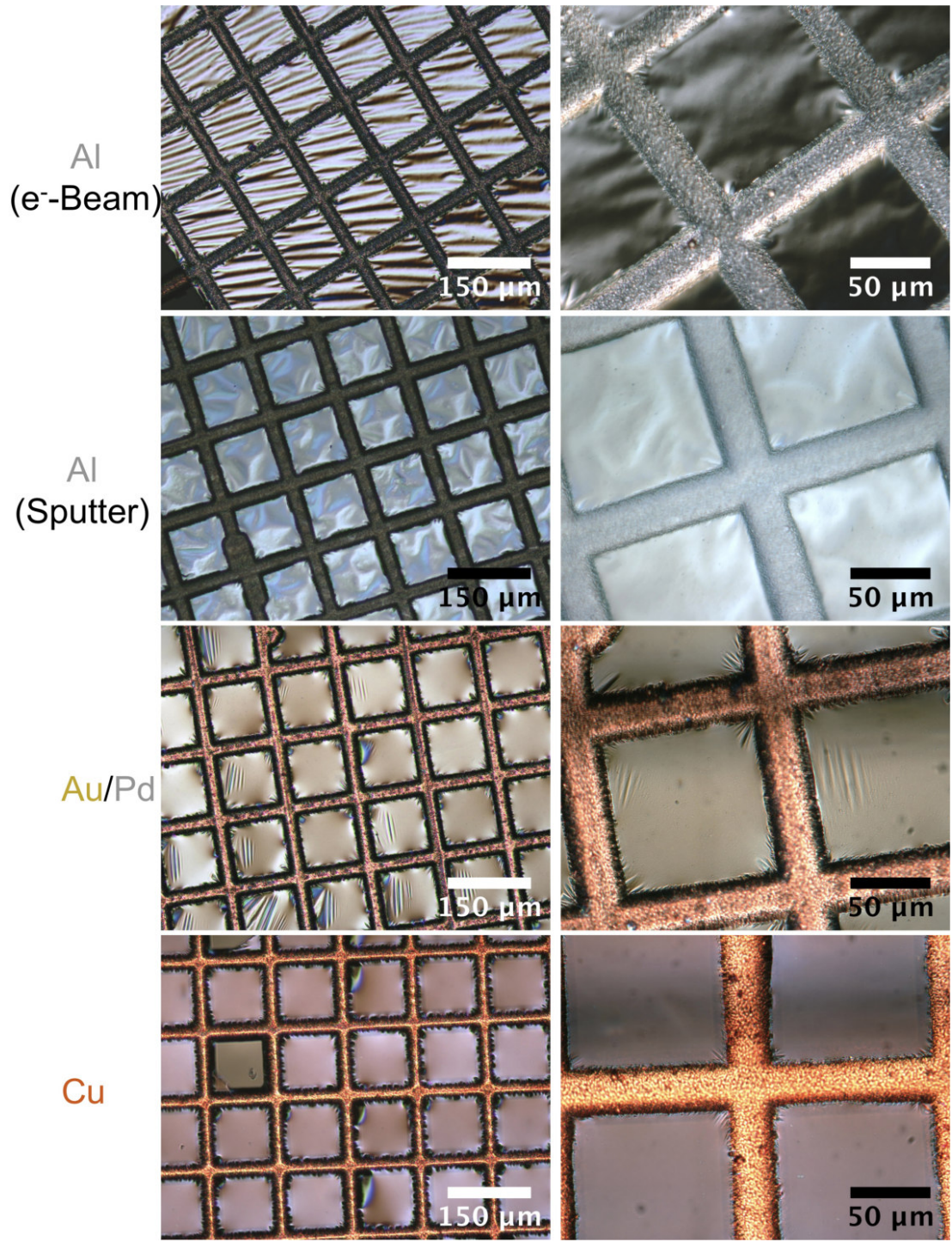

**Figure 3.19:** Light-microscope images (DIC) of all investigated metal-films (according to labeling) after removing the support-film. All films show a good macroscopic homogeneity and are rather smooth.

All films showed a good macroscopic homogeneity and were rather flat and smooth.

In general, thin-films are rather sensitive. A nagitation due to careless handling, can leave thin-films in most of the windows of one TEM-grid wrinkled and wavy. However, due to the high number of windows, almost always some of them with smooth thin-films can be found on a TEM-grid.

In summary, the technique presented here is very powerful and easy to execute. Self-sustaining thin-films of every material that can be applied by physical deposition-techniques and that is not etched by an $Ar/O_2$-plasma should be possible to make via this technique.

### 3.2.2 EELS-Measurements

EELS is capable of detecting single elements in a concentration below one atomic-percent combined with the high lateral resolution of TEM. In modern TEM, inelastically scattered electrons can be energetically separated by a magnetic prism. The energetically separated beam can be recorded by the CCD-camera of the microscope, yielding a spectrum with intensity as a function of the energy-loss of electrons.

When high energetic electrons are scattered inelastically in a specimen, different types of excitations are involved. The most important excitations are that of: plasmons (excitation of electrons in the conduction-band), valence-electrons, inner-shell-electrons and phonons (collective lattice-vibrations). At the excitation of plasmons, electrons are excited within the conduction-band, while valence- or inner-shell electrons are excited into higher lying energy-states, or into continuum. The excitation of phonons is equal to a heating of the sample. Those excitations have strongly different cross-sections, resulting in intensities that differ about many orders of magnitude for different excitations. The shape of a typical EEL-spectrum, is shown in figure 3.20. The zero loss peak contains the elastically scattered electrons. They are the main part of the recorded intensity (note that the intensity-scale is logarithmic). The plasmon-excitations follow in the region from $10\ldots30\,eV$. It is also called low-energy-loss region. Peaks corresponding to the excitation and ionization of valence- or inner-shell electrons follow in the high-energy-loss region. They are called ionization-edges, and lie upon an exponentially falling off underground. The ionization-edges have specific energies, that can be assigned to certain elements. The notation of the ionization-edges derives from the element and the shell from which the electron was excited. The signal corresponding to an excitation from the K-shell of an oxygen-atom is denoted as K-edge of oxygen. The term: $L_{2,3}$ denotes that electrons were excited from 2p-orbitals.

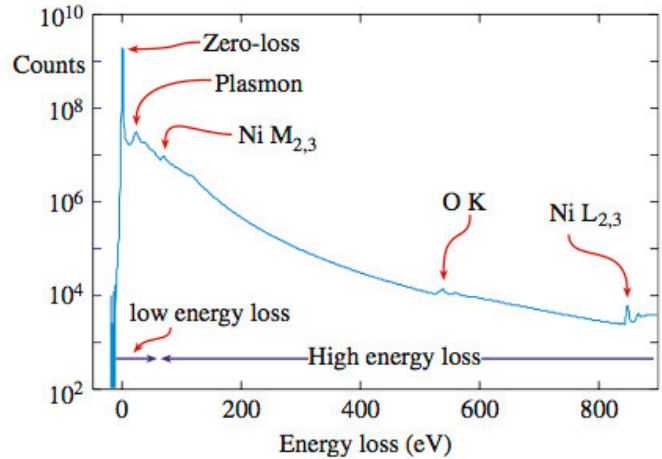

**Figure 3.20:** Typical EEL-Spectrum with logarithmic intensity-scale. The zero-loss peak is one to two orders of magnitude more intense than the plasmon peak, which in turn is many orders of magnitude more intense than the other features. On the tail of the plasmon-peak, L- and M-edges of Ni, and the K-edge of oxygen is visible [36].

The background of the spectrum is usually subtracted by fitting an exponential fall-off-function to a region of the background, lying before the corresponding signal. The integrated intensity, of the ionization-edges can be used, to quantify elements relative to each other, with the use of sensitivity-coefficients.

Figure 3.21 shows an EEL-spectrum of the background-subtracted K-edges for aluminium and carbon, measured on an Al thin-film.

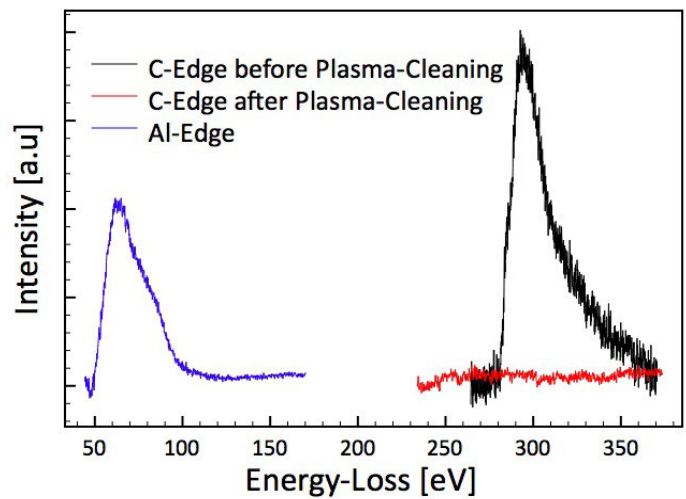

**Figure 3.21:** EEL-spectrum of an Al-film. The Al- and C-edges are clearly visible. After 15 min of plasma-cleaning, the C-edge is completely removed.

Prior to Plasma cleaning, the carbon-edge is clearly visible. After 15 min of plasma-cleaning, no carbon-signal can be detected any more. Accordingly, the successful preparation of a self-sustaining, carbon-free metal-film was proven.

Additionally, the thickness of a sample can be determined using EELS. For this, the integrals of the elastically scattered part (zero-loss) and the inelastically part (normally plasmon-loss) have to be determined. This is done by fitting an exponential fall-off-function to the zero-loss peak, and subtracting all signals from inelastically scattered electrons, yielding the needed integrals. With knowledge of the effective element-number of the specimen, its thickness can be calculated from the quotient of the obtained integrals.

An EEL-spectrum of the zero- and low-loss (inset shows magnified plasmon-loss) region of an Al-film in shown figure 3.22.

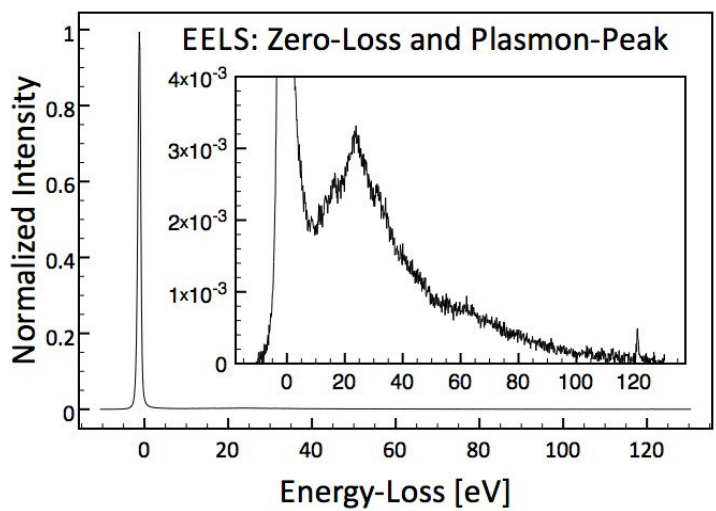

**Figure 3.22:** EEL-spectrum of an Al-film, showing the zero-loss and the magnified plasmon peak (inset). The spectrum was used to calculate the thickness of the film (see subsection 3.2.1).

The thickness value obtained with EELS was $19.66\,nm$.

### 3.2.3 Phase-Plate Preparation

One of the Al-films deposited via $e^-$-beam evaporation was tested as phase-plate. Further testing using other metal films, could not be done within the scope of this work, due to a prolonged downtime of the FIB-system.

The film-thickness of the tested phase-plat was determined to $17.95\pm0.343\,nm$ by AFM (see table 3.1.2). This thickness is in good correlation with the calculated value for a $-\pi/2$ phase-shift which is 17.38 nm at 200 kV. The film thickness was also determined by EELS. Her the measured value was 19.66 nm. This value is 8.7% higher compared to the AFM data. A secure judgment about which result is more reliable can not be given, due to the absence of further EELS-measurements. In the literature, the accuracy of EELS for thickness-measurements is estimated to be better than $\pm 20\%$ [31]. The accuracy of AFM is estimated to be better than $\pm 10\%$. Therefore the expected film-thickness and phase-shift should lie in the range of $18\ldots18.5\,nm$ and $-(0.52\pi\ldots0.53\pi)$ respectively.

The central hole was milled into the phase-plate from the bottom-side, with the carbon-film facing towards the ion-beam. In this way, the carbon-film was supposed to work as a shield against misdirected $Ga^+$-ions. Figure 3.23 shows SEM-images of the Al-film after milling the central hole. a.) shows an overview of the Al-film. The whole film looks homogeneous and smooth. Also at higher magnifications (panel b.) and c.)), no microstructure is visible. Panel b shows the position of the hole. It can be seen that the hole is off center. However, this should not diminish the properties of the phase-plate much, because the distances to the edges of the grid should still be larger than $30\,\mu m$. Accordingly, the theoretical lower limit for phase-contrast transfer should not be higher than $0.25\,nm$ (compare equation 2.13), which is still low enough to image small biological structures. In fact, the de-centered hole is unlikely to restrict the resolution significantly; diminishing influences, such as the periodicity of the $CTF$ are larger than that of the de-centered hole. Panel b.) shows a detailed view of the central hole. The hole has a diameter of $1.17\,\mu m$. This is 15% larger than the planned value and will increase the cut-on frequency from originally $1/15.06\,nm^{-1}$ to $1/13.10\,nm^{-1}$. However, this should still be enough for phase-plate tests.

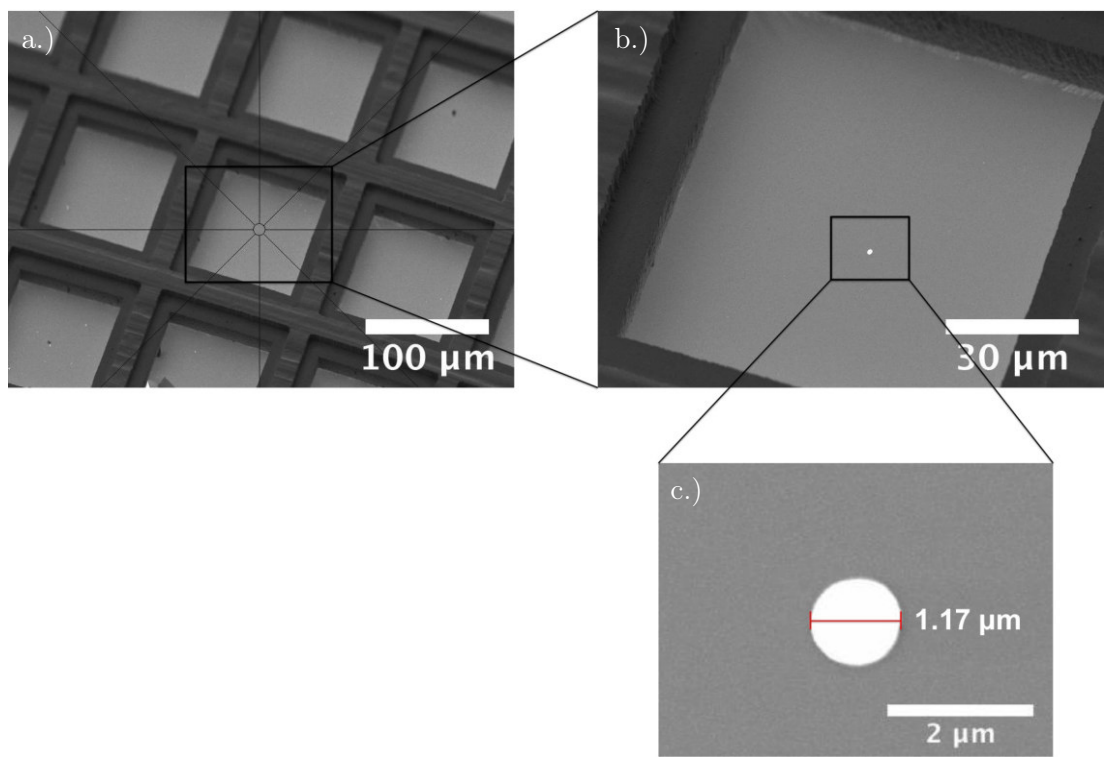

**Figure 3.23:** SEM-images of an aluminium phase-plate taken with a high-tension of $5\,kV$. a.) shows an overview of the TEM-grid from the backside of the film. The film was mounted with the carbon support-film facing towards the ion-gun. In this way, the carbon-film might work as a protecting-shield for the Al-film against redeposition. b.) Window of the TEM-grid with FIB-milled central hole. It can be seen, that the hole is not ideally centered in the TEM-window. This should not however restrict the properties as a phase-plate (see text below). The size of the central hole (c.) determined with SEM is $1.17\,\mu m$ which is bigger than the nominal value of $1\,\mu m$.

## 3.2.4 Application as Phase-Plate

The Al-films were tested as phase-plates following the same procedure as described in subsection 3.1.4. Images were taken using the same microscope (Zeiss Libra 200) and under similar conditions as with the carbon phase-plate.

The effect of the phase-plate is illustrated in figure 3.24. Like for the carbon phase-plate, a big change in contrast is visible in the TEM-micrographs (panel a.) and b.). In panel c.), a negative phase-shift, indicated by red arrows is visible in the corresponding FFTs. The phase-shift is also visible in the power-spectra in panel d.).

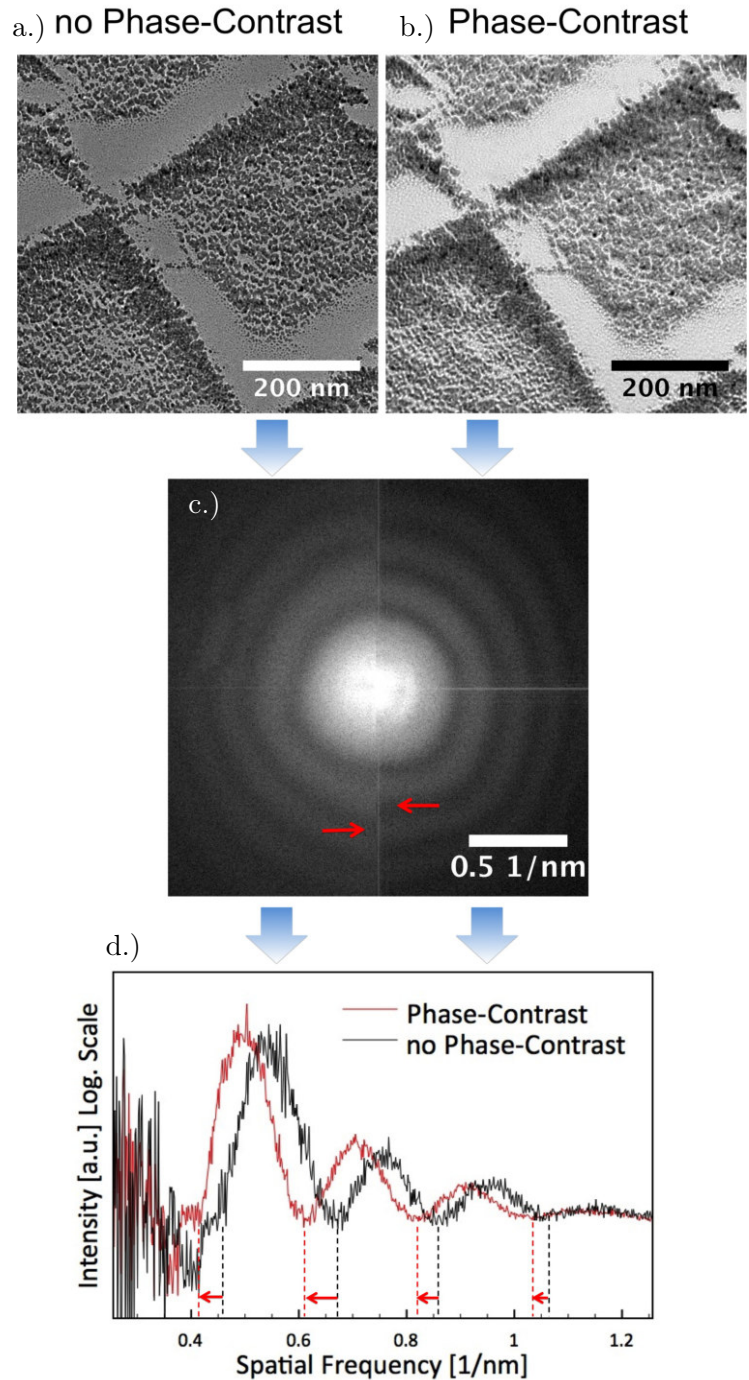

**Figure 3.24:** Panels a.) and b.) show TEM-micrographs of a reference-sample with the same high underfocus. a.) was taken with the phase-plate in the back-focal-plane but not with the central hole on the optical axis, and b.) with the centered phase-plate. Panel c.) shows the FFT of a.) on the left and of b.) on the right side. Red arrows mark the displacement of the thon-rings. d.) shows scans of the FFTs averaged over 90° of rotation. Dashed-lines mark minima of the power-spectra, which are clearly apart from each other.

Simulated $\left|CTF^2\right|$ were tried to fit to the obtained power-spectra. When the phase-plate was retracted from the back-focal-plane, the simulated $\left|CTF^2\right|$ matches the zero-points of the power-spectrum (not shown here), delivering a defocus-value of $572\,nm$. Figure 3.25 shows power-spectra for the aligned and unaligned phase-plate in the back-focal-plane with a logarithmic intensity-scale.

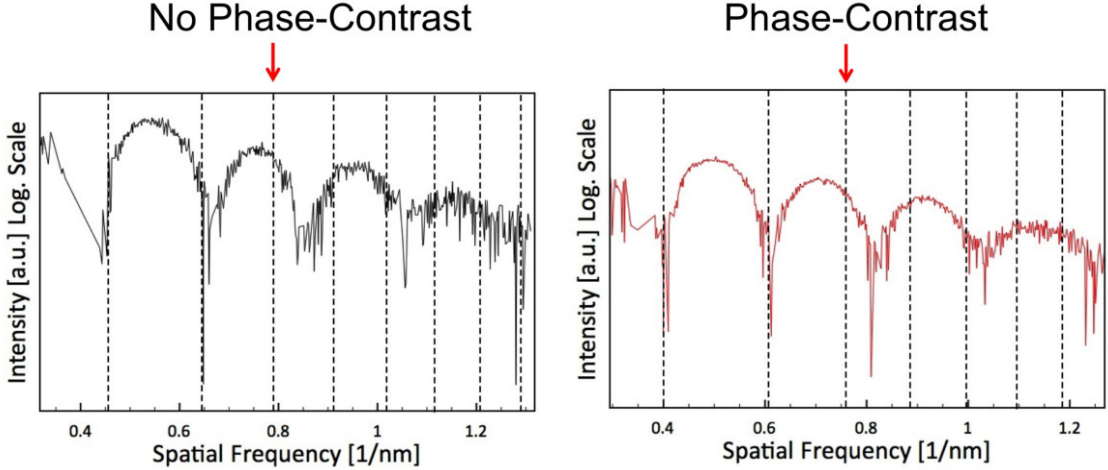

**Figure 3.25:** Power-spectra of TEM-micrographs, obtained with an unaligned (left) and aligned (right) phase-plate in the back-focal-plane. The intensity-scale is logarithmic so that minima are better visible. The minima of a calculated squared modulus of the CTF (dashed lines) can not be fitted fairly to both spectra. The parameters for the CTF calculations were: spherical-aberration: $1.2\,mm$; electron-energy: $200\,keV$.

For both the unaligned (left panel) and the aligned (right panel) phase-plate in back-focal-plane, the power-spectra do not match the zero-points of the calculated $\left|CTF^2\right|$ (dashed lines). In the left panel, zero-points were calculated with a defocus-value of $1900\,nm$. For the rigth panel, the same defocus plus a constant phase-shift of approx. $-\pi/4$ was chosen. Both spectra show similar deviations from the calculated zero-points.

If the "phase-contrast" power-spectrum is manually shifted by $\pi/4$, the two spectra are almost congruent. They are shown in figure 3.26 displaced vertically for clarity.

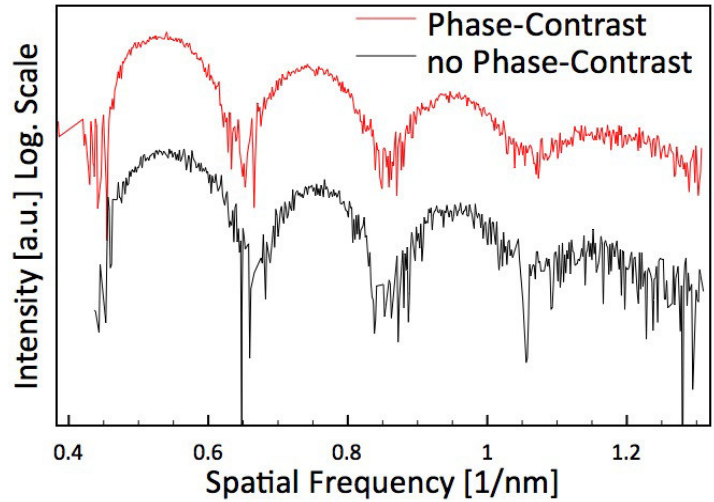

**Figure 3.26:** Power-spectra for an Al phase-plate shown in figure 3.24 displaced vertically for clarity. Intensity-scale is displayed logarithmic. The "phase-contrast" spectrum was shifted manually by approx. $\pi/4$. The two spectra are almost congruent.

The congruency of the two spectra indicates, that the difference between the two conditions is negligible, except for the constant phase-contrast phase-shift. Apparently, the thin-film of the phase-plate inserted in the back-focal-plane, is a suitable "no phase-contrast" scenario in the case of the Al phase-plate. For the carbon phase-plate, the "no phase-contrast" spectrum showed no congruency with the manually shifted "phase-contrast" spectrum. This is shown in figure 3.27.

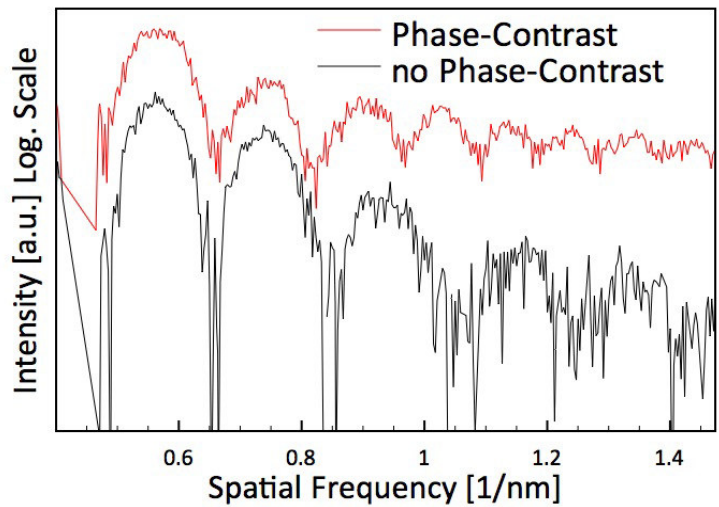

**Figure 3.27:** Power-spectra for a carbon phase-plate shown in figure 3.13 displaced vertically for clarity. Intensity-scale is displayed logarithmic. The "phase-contrast" spectrum was shifted manually. The two spectra are not congruent.

When the phase-plate is centered on the optical axis, only approx. 1% of the beam-current irradiates the phase-plate. This should decrease charging greatly for the centered phase-plate, compared to the uncentered phase-plate where most of the beam irradiates the film. Thus, the corresponding power-spectra should not differ only by a constant phase-shift if charging would occur. This indicates, that charging of the Al phase-plate is significantly reduced with respect to the carbon phase-plate or eliminated completely.

A further indication for a reduced charging of the Al phase-plate, is that the FFT of the TEM-micrograph did not change noticeably over time in contrast to the carbon phase-plate. Both differences of congruency of the power-spectra and of the changing FFT with time indicate a greatly decreased charging of the Al phase-plate due to the absence of adsorbed gases and a higher conductance with respect to carbon.

Although charging is probably reduced or avoided completely, there is still a distortion in form of a spatial-frequency dependent phase-shift applied to the power-spectrum when the Al thin-film is inserted into the back-focal-plane (see figure 3.25). Another reason for the distortion might be an inhomogeneous thin-film phase-plate.

It is surprising, that the phase-shift induced by the Al phase-plate is less than 50% of the expected value. The uncertainty of the mean inner-potential value should not be bigger than 25% while the thickness of the film was determined with relative high accuracy. In all likelihood, there should be other effects decreasing the phase-shift induced by the Al phase-plate.

For the Al phase-plate no beam damage was observed after several hours of beam-exposition with imaging the phase-plate using the combined spot-diffraction-mode. Thus the durability of the Al phase-plate should be high as was expected for a metal.

To find out more about the distortion of the power-spectrum, metal-films were examined with TEM.

## 3.2.5 Characterization with TEM

The micro- and nanostructure of all fabricated metal-films was analyzed using transmission-electron-microscopy. Both, bright-field TEM-micrographs and SAD-pattern were recorded. A JEOL JEM2200 FS was used for all investigations except of the Al-film made by e$^-$-beam evaporation. Here, a ZEISS Libra 200 (CRISP) TEM was used for TEM-micrographs (both instruments were operated at 200 kV).

SAD-pattern were recorded using a camera-length of $30\,cm$ and a small diffraction-aperture. Sample-areas were chosen, that contained nano-crystallites and parts of the flat film, to best represent the average of the whole film. Results are given in figure 3.28.

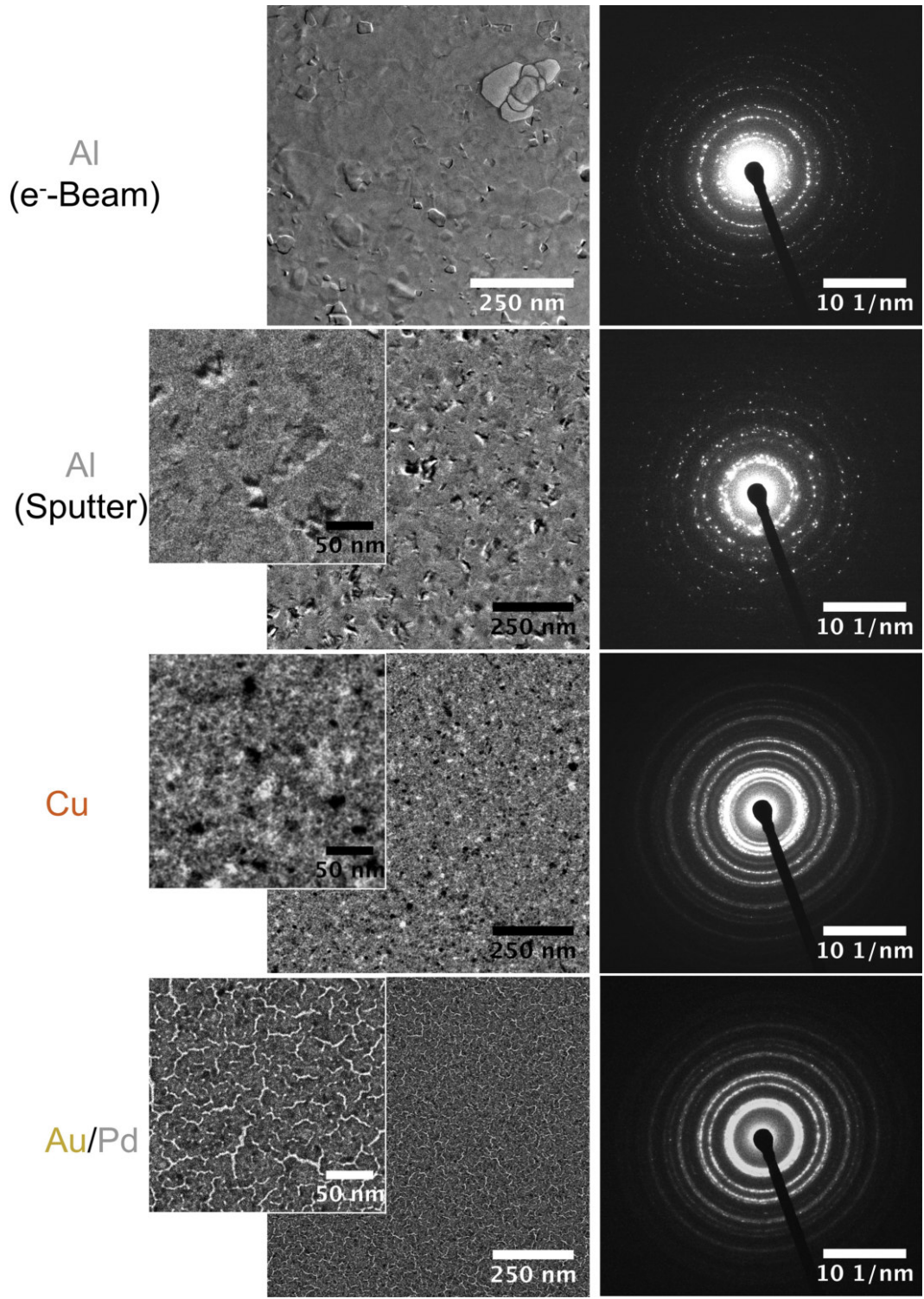

**Figure 3.28:** TEM-micrographs (left) and selected-area-diffraction images (right) of all different metal-films according to the labeling. For TEM-micrographs of the sputtered Al- Cu- and Au/Pd-films the insets show magnified cut outs of the images. All metal films exhibit a nanometer-sized film-structure. For the Al-films, nano-crystallites are clearly visible causing single diffraction-spots in the corresponding SAD-images. For the Au/Pd- and the Cu-film, the crystalline domaines are smaller resulting in nearly continuous diffraction-rings. All films are polycrystalline.

All different metal-films are polycrystalline and show nano-crystallites of different sizes. Accordingly, SAD-images show relatively sharp diffraction-rings and even diffraction-spots. The Au/Pd-film has no continuous surface, but has gaps corresponding to the strong tendency for island-growth of gold. The Al-films show bigger nano-crystallites than other films. In case of the Al-film deposited by $e^-$-beam-evaporation, crystallites have sizes ranging from $100 \ldots 200\,nm$. Crystallites of this size are likely to distort the power-spectrum noticeable, having a size of $10 \ldots 20\%$ of the central hole. In general, polycrystalline thin-films are less homogeneous than completely amorphous films, as can be seen in the TEM-micrographs, especially when compared to the carbon-film. A high crystallinity is also unfavoured, because of increased electron-scattering due to diffraction.

## 3.2.6 Discussion and Outlook

The method for self-sustaining metal-film preparation of Janbroers et al. could be transferred to other metals. In agreement with his reports, the carbon-film was removed selectively after approx. 15 minutes of plasma-cleaning with an Ar/O$_2$-Plasma. This was verified by EELS. After the carbon-film had been removed, films were rather smooth.

As was expected, there is a change of contrast visible in TEM-micrographs, and a negative phase-shift in the corresponding power-spectra induced by the phase-plate. Surprisingly, the phase-shift is approx. $-\pi/4$, which is below 50% of the calculated value.

Evaluation of the obtained TEM-micrographs indicate, that charging is decreased for the Al phase-plate with respect to the carbon phase-plate or eliminated completely, as expected by the properties of the materials. Leaving the continuous Al thin-film in the back-focal-plane, seems to be a suitable "no phase-contrast" scenario. However, there is still a distortion of the power-spectrum induced by the Al thin-film.

Electron diffraction revealed, that the fabricated metal films are crystalline. The nano-crystallites could be a reason for the distortion of the power-spectrum. In order to reduce their crystallinity, the preparation of metal-films was changed which will be presented in the next section.

## 3.3 Cryo-Deposited Metal-Films

As we have seen in the last subsection, metal-films prepared at room-temperature (RT), possess a certain crystallinity. Probably the easiest way to reduce the crystallinity of a solid, is to lower the temperature during formation. In case of a liquid, its low ordered structure is best preserved into the solid-state, when it is quenched quickly. A low temperature withdraws activation-energy from the reacting-particles, that is needed for the process of crystallization. The same principle can be adopted to the preparation of a thin-film with vapor-deposition-techniques. Here, the reacting particles are atoms or small clusters consisting of several atoms that are in the gas-phase. When particles from the gas-phase hit the surface of the substrate, they condense. Due to their kinetic-energy and the lattice-energy that is won by condensation, the particles are able to diffuse on the surface and hence enter lattice-sites that are energetically more favorable. An ordered crystal-lattice is formed as a result. By quenching the incoming particles, this process is prevented, and the particles are rather randomly distributed in the formed solid.

### 3.3.1 Preparation

All cryo-deposited metal-films (cryo-films) were produced like metal-films described in 3.2.1, but with a lower temperature during deposition. This was done using the MED020 (BAL-TEC AG) combined with a "VCT 100 Vacuum-Cryo-Transfer-System" (BAL-TEC AG). Here copper-parts inside the vacuum-chamber were put in thermal contact with a liquid nitrogen reservoir. Thus, the pressure could be lowered to $1 \cdot 10^{-7}\,mbar$. After this pressure was reached, samples, mounted on a copper block, were inserted into the recipient of the MED020 using a manipulator. The temperature of the samples was gradually lowered to $-147\,°C$ in the vacuum.

Cryo-deposited Al- and Au/Pd-films were produced via sputtering. Deposition-parameters were kept the same as for the corresponding room-temperature metal-films except for the temperature. Figure 3.29 shows light-microscope images of Al- and Au/Pd-cryo-films.

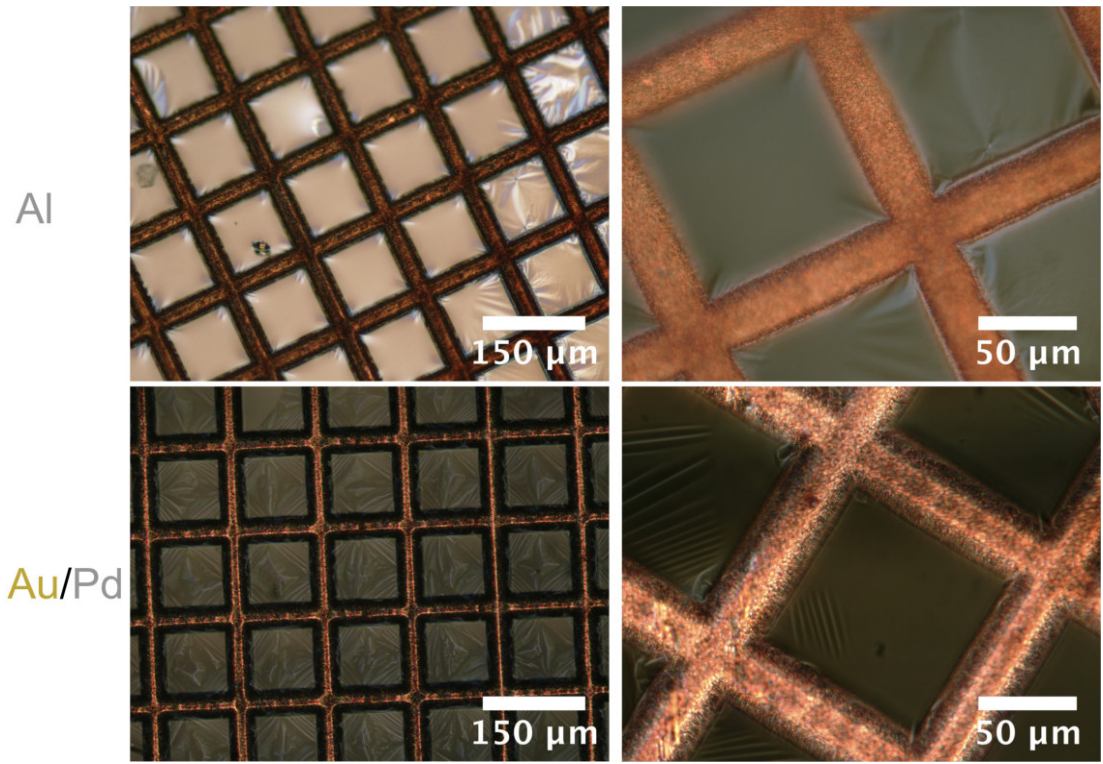

**Figure 3.29:** Light-microscope images (DIC) of all cryo-metal-films. Corresponding to a lower crystallinity and a lower electric-conductivity, the cryo-films are transparent.

In contrast to metal-films deposited at RT, cryo-films are very transparent, even though they have similar thicknesses. The transparency indicates, that the electric-conductivity is lower than of the RT-metal-films. Usually all metals are opaque to visible light, because of their high electric conductivity. Through the broad band of energy-levels available to the Fermi-gas, all wavelengths of the visible-light are absorbed. The observation of the higher transparency indicating a lower electric-conductivity, would also be in agreement with a lower crystallinity. If a solid is amorphous rather than crystalline, atomic-orbitals can not form broad, overlapping energy-bands like in crystalline metal, thus visible light is not absorbed. As we mentioned in the introduction to section 3.2, the same is true for carbon and graphite. While amorphous carbon exhibits poor conductivity, crystalline graphite is a good conductor.

Figure 3.30 shows magnified light-microscope images of a cryo Al-film.

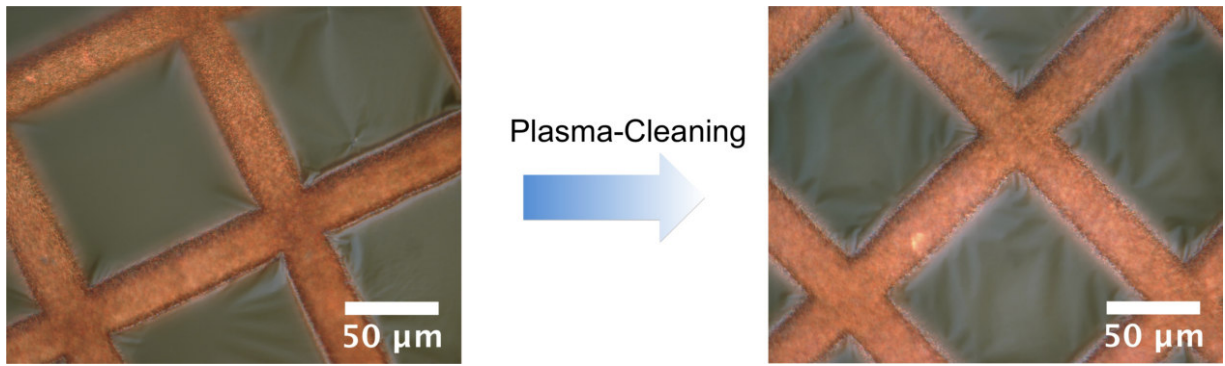

**Figure 3.30:** Light-microscope images (DIC) of an aluminium cryo metal-film. No big affect on the flatness of cryo-films through plasma-cleaning can be observed.

No big difference concerning the flatness of the film before and after plasma-cleaning could be observed with the light-microscope. The films behave mechanically similar to the RT-deposited.

## 3.3.2 Characterization with TEM

Cryo-films were characterized by TEM equally to other-films. All images were taken on the JEOL JF2200 operated at $200\,kV$. The same imaging conditions as for the carbon and RT metal-films were chosen. Figure 3.31 show the corresponding TEM- and SAD-images.

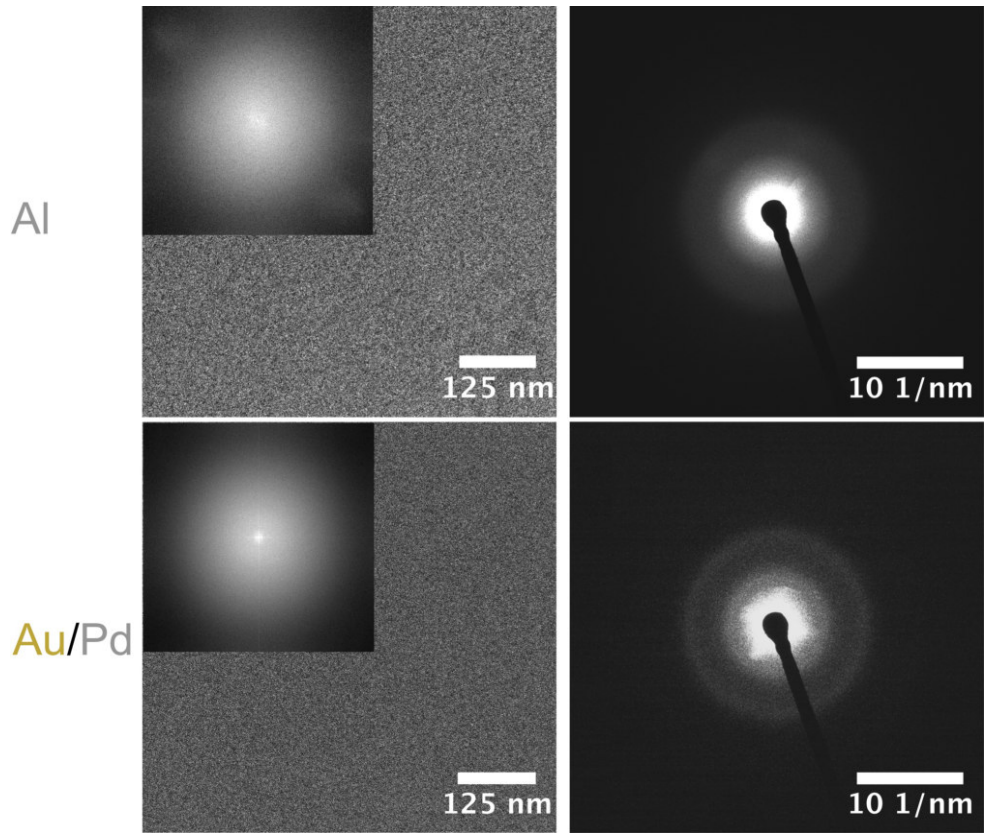

**Figure 3.31:** TEM-micrographs (left) and SAD-images (right) of all cryo-metal-films according to the labeling. Similar to the carbon-film (compare figure 3.14), cryo-films exhibit no crystalline structure. Insets showing FFTs, prove that the corresponding TEM-micrographs were taken at focus. SAD-images show that the cryo-films are highly amorphous.

All cryo-films exhibit no crystalline structure as can be seen in both TEM- and SAD-images. They are completely amorphous, as is shown by the broad diffraction-ring with low intensity in the SAD-images. The big difference in crystallinity, with respect to the RT-deposited films, becomes obvious in figure 3.32.

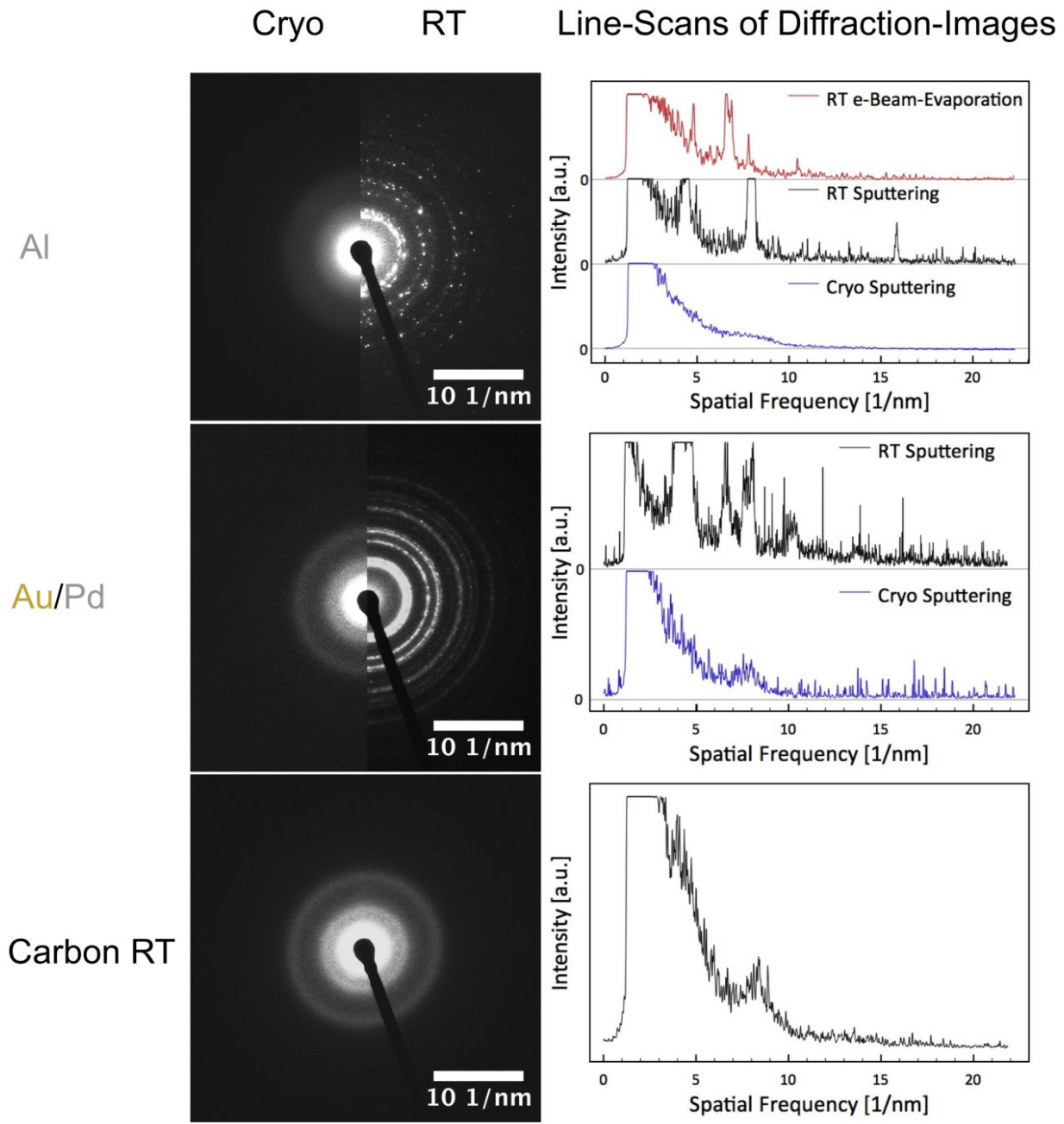

**Figure 3.32:** Comparison between room-temperature- and cryo-deposited films according to the labeling. The difference in crystallinity is clearly visible in the SAD-images. Again the difference is obvious in the corresponding line-scans shown on the right. The scan-curves are displaced vertically for clarity. While all RT-deposited films exhibit a diffraction-pattern, the curves of the cryo-films show essentially only one very broad reflex each. For comparison the SAD-image of the RT-deposited carbon-film is shown as well.

Here the SAD-images of the cryo-films are directly compared to the corresponding RT-deposited films, by showing one half of each SAD-image in one image. Additionally, the SAD-image of a carbon-film is shown again for comparison. It looks very similar to those of the cryo-films. On the right, line-scans of the corresponding SAD-images, are shown. While all RT-deposited metal-films show reflexes in the diffractograms, essentially only one very broad is observed in the diffractograms of the cryo-films and of the carbon-film.

Because metals tend to crystalize very readily even at ambient conditions, the crystallinity of the cryo-films was controlled again after a certain period of time at RT and atmospheric-conditions. Figure 3.33 shows the corresponding SAD-images. The SAD-image for the Al-film after $4 \ldots 6$ weeks was recorded with a LEO 922 OMEGA TEM with a $LaB_6$-cathode as electron-gun operated at $200\,kV$. All images were taken with a camera-length of $30\,cm$. The direct comparison between the SAD-images reveals no crystallization after a period of $4 \ldots 6$ weeks.

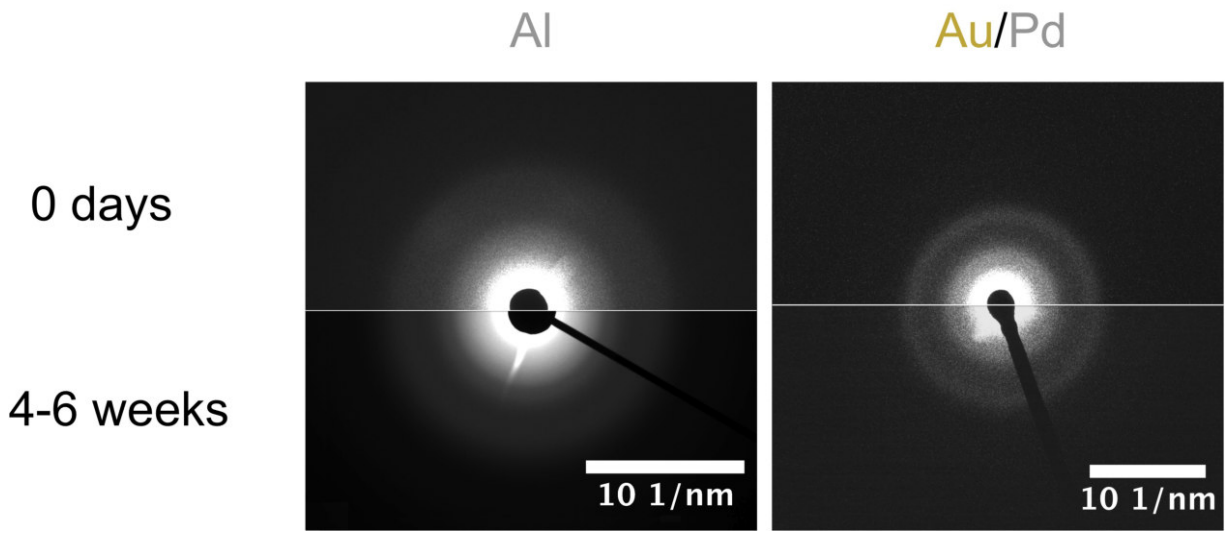

**Figure 3.33:** SAD-images of cryo-films corresponding to the labeling showing the influence of aging on the crystallinity. After $4 \ldots 6$ weeks at RT and atmospheric conditions, no crystallization can be observed.

Not only was the influence of time on the crystallinity of the cryo-films determined, but also of beam-exposition. For this, SAD-images taken after a certain time of electron-beam deposition, were scanned radially (zero at the center of the image), and were averaged over $360°$ of rotation. Figure 3.34 shows this procedure for the SAD-image after 0 min of beam-exposition on the left, and the according rotationally averaged scan on the right.

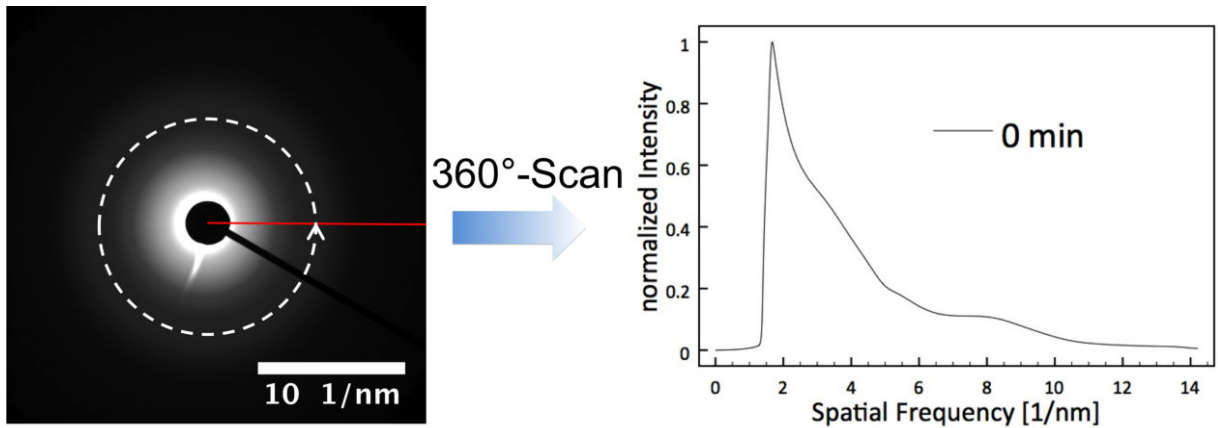

**Figure 3.34:** Rotational averaging of an SAD-image to yield a diffractogram. SAD-image after 0 min of beam-exposition (left) and the according scan (right).

All SAD-images were taken at a camera-length of $30\,cm$ and with the same imaging-conditions at a LEO 922 OMEGA TEM operated at $200\,kV$. The electron-dose during exposition measured on the fluorescent screen was 4300 electrons/$(nm^2 \cdot s)$. The brightness and contrast was adjusted equally for each image as well. d-spacing was calibrated, using the interlayer distance of Au(111) lattice-planes measured with a polycrystalline Au- sample, as reference.

The influence of beam exposition is shown in figure 3.35.

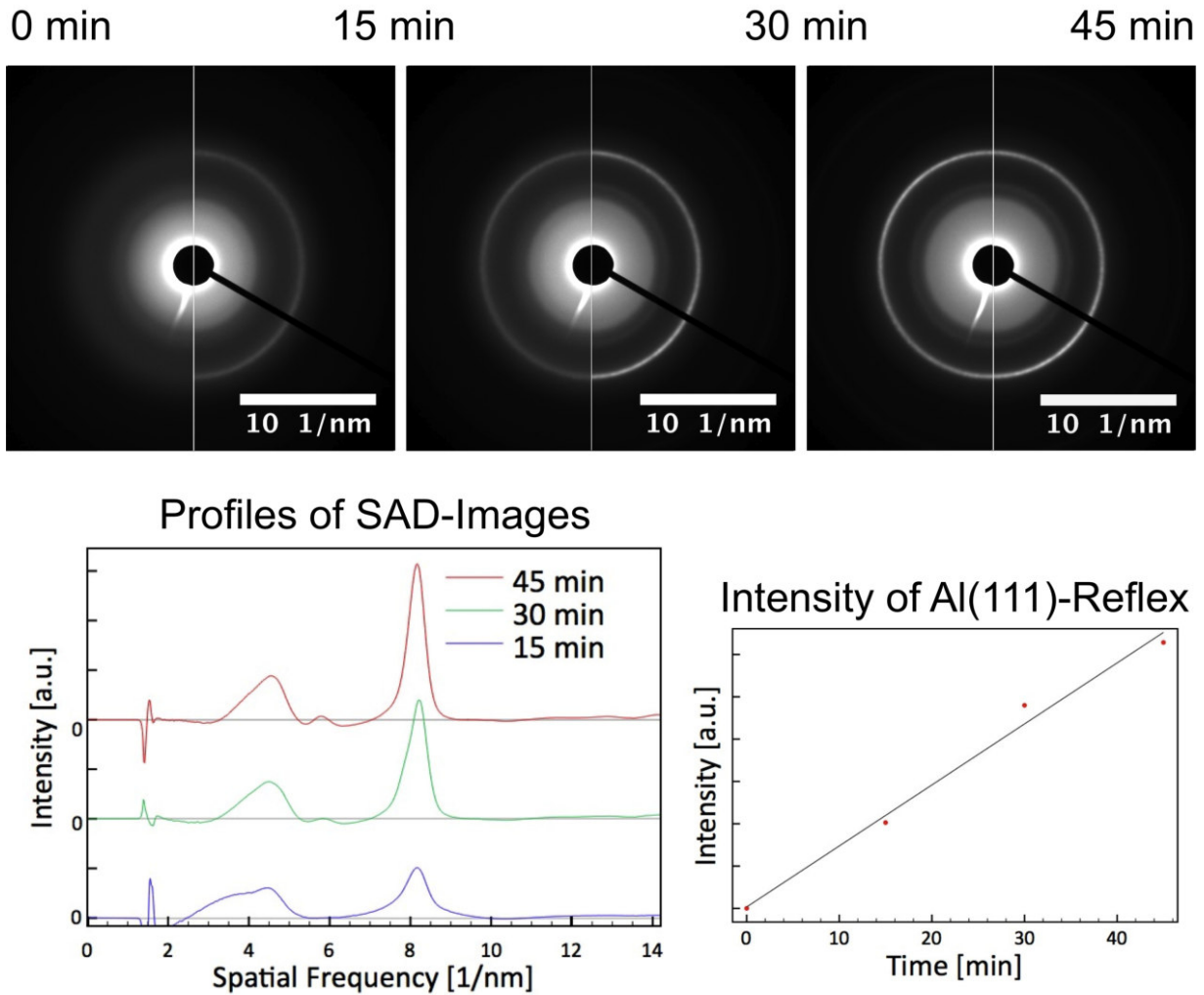

**Figure 3.35:** Comparison between SAD-image of an cryo-Al-film after 0, 15, 30 and 45 minutes of beam-exposition, according to the labeling. An increase of crystallinity with increasing beam-exposition can be observed by the formation of a relatively sharp diffraction-ring. The Beam intensity was kept at a high, constant level during the exposition. The "peak" around $1.7\,nm^{-1}$ in the profiles of SAD-images is an artifact due to bright halo of the covered central diffraction-spot. Rotationally averaged scans of the SAD-images show a linear increase of the Al(111)-reflex's intensity with time.

The comparison of SAD-images, recorded after increasing times of beam-exposition reveals an increasing of crystallinity. This can be seen by the formation of a relatively sharp diffraction-ring corresponding to the Al(111) lattice-plane. The diagram below shows rotationally averaged scans of the SAD-images. They are displaced vertically for clarity. Highest intensities were normalized to unity. To accentuate the change between subsequent diffractograms, the starting diffractogram (figure 3.34, right) was subtracted from the others. The "peak" around $1.7\,nm^{-1}$ in the profiles of SAD-images is an artifact due to bright halo of the covered central diffraction-spot. The intensity of the (111)-reflex increases linearly with time (right). After 45 minutes exposition to a high-current beam, the crystallinity is still much lower compared to the RT-deposited Al-films. Further investigations about the influence of the electron-dose applied to the phase-plate at the back-focal-plane have to be carried out, to be able to say how the crystallinity is changed there and how the beam-induced crystallization lowers the properties of the phase-plate.

### 3.3.3 Discussion and Outlook

The deposition at about $-147\,^\circ C$ yielded amorphous metal-films, as was shown by electron-diffraction. According to a low crystallinity and in turn, to a lower conductivity, cryo-metal films are transparent. Those cryo-deposited films, could be an ideal material for Zernike phase-plates, due to there low crystallinity and high homogeneity. While they showed no recrystallization over a certain period of time at RT and atmospheric conditions, this was the case after a prolonged exposition to a high-current beam. However, even after applying a very high electron-dose to the film, the crystallinity was still much lower compared to the RT-deposited metal-films. Tests of cryo-Al-films as phase-plate are planned. Due to technical difficulties, they could not be tested so far.

# 4 Conclusion

Ten years after the Zernike phase-plate was implemented successfully for TEM, another material than carbon was tested here for the first time. Thin-films were produced according to Danev et al. and Janbroers et al. The technique by Janbroers et al. proved to be valuable for thin-film preparation of metals. A carbon and an aluminium thin-film were processed into Zernike phase-plates, by milling a central hole with a diameter of $1\,\mu m$ into them, using a focused-ion-beam. The thickness of thin-films was measured using AFM. Step-edges of those films produced on mica-substrates were well suited for this purpose.

In the past, Danev et al. tried to quantify the phase-shift induced by a Zernike phase-plate, by comparing TEM-micrographs taken without phase-plate and with the aligned phase-plate at the back-focal-plane. For quantification of the mere phase-contrast phase-shift, this procedure is inadequate, because of the thin-film's influence on the image. Instead, it was tried to quantify the phase-shift here, by comparing micrographs taken with the continuous thin-film of the phase-plate and with the aligned phase-plate at the back-focal-plane. Although the phase-shift obtained here had the wrong sign, the continuous thin-film at the back-focal-plane is considered a more suitable "no phase-contrast" scenario with respect to the retracted phase-plate.

The squared modulus of the $CTF$ or the phase-contrast $CTF$, respectively, could not be fitted satisfactorily to the power-spectra of these micrographs even when a change of defocus (lens-effect of the thin-film) was included. The $CTF$ has to be modified when a Zernike phase-plate is inserted into the back-focal-plane, in order to account for the distortion due to the thin-film. This distortion is assumed to be a spatial-frequency dependent phase-shift. The reason for this distortion is most likely charging of the carbon phase-plate itself or of adsorbed particles. To determine these distortions more precisely, further investigations have to be made.

For the Al phase-plate, a different distortion was observed. Although $\left|CTF^2\right|$ could not be fitted satisfactorily to the power-spectra for aligned and unaligned phase-plate either, here power-spectra are almost congruent when one is shifted manually by a constant phase-shift over the other. This indicates that for the Al phase-plate, the two conditions are adequate to quantify the phase-shift and that no charging seems to occur. The Al phase-plate exhibits a good mechanical stability and constant imaging conditions while used. Using the carbon phase-plate the image's FFT changed rapidly with time shortly after insertion of the phase-plate. This further supports the assumption that no charging of the Al phase-plate occurs. However, the Al phase-plate still causes a distortion in power-spectra of images. The high crystallinity of the thin-film is likely to be at least partly responsible for this distortion, because it reduces the homogeneity of the thin-film and increases scattering. To avoid a high crystallinity, metal-films were deposited under cryo-conditions. Those cryo-films were completely amorphous. They showed no or only low recrystallization after a long period of time under atmospheric conditions or under exposure to a high-current electron-beam, respectively. Due to technical difficulties, tests with these films as phase-plates have not been done yet, but are planned for the future. Cryo-films are a very promising material for Zernike phase-plates.

The phase-shift for the Al phase-plate was negative as expected, but only approximately half of the expected value. The uncertainties of the thickness of the film, or of the mean inner-potential are not so large as to account for this high deviation. Further experiments have to be carried out, to point out reasons for it. A comprehensive knowledge about the influence of the phase-plate's thin-film on the obtained image is important, to avoid distortions and thus improve the quality of phase-contrast by the phase-plate.

The experiments carried out during this thesis, opened promising perspectives of thin-film phase-plates. Problems concerning the preparation and the thickness-measurement of phase-plates could be solved. This lays the basis for future experiments on the quantitative phase-shift behavior, and the optimisation of these phase-plates.

# 5 Danksagung

Mein erster Dank gilt in jedem Fall Herrn Prof. K. Wandelt. Ohne seine Unterstützung wäre diese Diplomarbeit nicht möglich gewesen. Schon während meines Studiums hat er mir des öfteren geholfen. So hat er mir die Möglichkeit geschaffen mein Nebenfach Oberflächenchemie zu einer für mich günstigen Zeit zu absolvieren. Er hat bei mir ein großes Faible für Oberflächenwissenschaften geweckt und meine Orientierung als Wissenschaftler nachhaltig beeinflusst. Auch zuletzt während meiner Diplomarbeit hat er mich als mein Erstgutachter sehr gut betreut und meiner Diplomarbeit dank seiner sehr guten Ideen (Cryo-deposition!), neuen Zündstoff geliefert. Es ist mir eine groe Freude, das ich meine Diplomarbeit noch bei ihm anfertigen konnte. Ich schätze ihn als Menschen und als Wissenschaftler und werde nie vergessen das er sich (trotz widriger Umstände) meiner angenommen hat. Hierfür vielen, vielen Dank!

Mein Dank gilt natürlich auch Herrn Prof. U. Kubitschek ohne dessen Unterstützung meine Diplomarbeit ebenfalls nicht möglich gewesen wäre.

Der Stiftung Caesar an deren Forschungszentrum alle Arbeiten für meine Diplomarbeit durchgeführt wurden, bin ich zu großem Dank verpflichtet. Durch die exzellente Ausstattung wurden mir sehr gute Rahmenbedingungen für meine Diplomarbeit geschaffen.

Mein Betreuer Dr. Stephan Irsen hat mich sehr gut betreut. Obwohl von ihm geäußerte relative zeitliche Angaben, bezüglich eines bestimmten Ereignisses ("Ich bin gleich bei dir.", Ich komme in 10 Minuten zu dir in den Bunker.") von den gesellschaftlichen Konventionen und der wissenschaftlichen Zeitdefinition mitunter um einen Faktor 1000 abwichen, war sein unermüdlicher Einsatz, Garant für das gelingen meiner Arbeiten. Die Arbeit mit diesem Allroundtalent hat mir viel Spass gemacht und die zahlreichen Diskussionen mit ihm waren überaus fruchtbar. Vielen Dank!

Die Gruppe Elektronenmikroskopie und Analytik hat mir ein sehr angenehmes und gutes Arbeitsumfeld bereitet. Angelika Sehrbrock bin ich sehr dankbar für das Lochen der Phasenplatten. Dem KonTEM-Team: Steffen Pattai, Patrick Kurth und Joerg Wamser bin ich dankbar für dessen Unterstützung bei Phasenplattentests. Johannes Lenz, Heiko Linnenbank, Felix von Cube, Dr. Jürgen Feydt und Kristina Bachmeier bin ich auerdem sehr dankbar für anregende Diskussionen, Tips oder Korrekturen an meiner Diplomarbeit. Allen anderen Mitgliedern der Gruppe danke ich außerdem für die gute und freundliche Atmosphäre.

Ein wichtiges Kapitel meines Lebens geht nun zu Ende. Ein guter Zeitpunkt um anderen Personen zu danken die nur indirekt an der Entstehung meiner Diplomarbeit beteiligt waren.

Prof. R. Glaum hat mich während meines Studiums begleitet unterstützt, und mich als

Wissenschaftler beeinflusst. Seine Ratschläge waren sehr hilfreich. Für sein Engagement über die Jahre bin ich ihm sehr dankbar.

Dr. Wilfried Assenmacher ist mir ebenfalls in sehr guter Erinnerung geblieben. Das Seminar zum anorganisch-chemischen Fortgeschrittenenpraktikum war wirklich eine sehr gute Veranstaltung. Mein Interesse für Festkörperchemie/physik hat sich insbesondere dort entwickelt.

Die Professoren Vöhringer und Gansäuer waren massgeblich an meiner Ausbildung in PC und OC beteiligt. Besonders Prof. Vöhringer danke ich (jetzt wo ich alles hinter mir habe) für seine "harte" Schule.

Meiner Familie bin ich ebenfalls sehr dankbar (besonders meiner Mutter natürlich).

An meine Lieben: Wem ich am allermeisten dankbar bin, brauche ich hier nicht zu erwähnen. Ihr wisst es auch so. Ich weiss nicht genau wie diese Sache ohne euch ausgegangen wäre.

Diese Arbeit ist für euch!

# List of Figures

# List of Tables